\documentclass{aa}  
\usepackage{graphicx}
\usepackage{txfonts}

\begin{document}
   \title{Polycyclic aromatic hydrocarbon selected galaxies
   }
   \author{Martin Haas
          \inst{1}
          \and
          Christian Leipski
          \inst{2}
          \and
          Ralf Siebenmorgen
          \inst{3}
          \and
          Helmut Meusinger
          \inst{4}
          \and
          Holger Drass
          \inst{1}
          \and
          Rolf Chini
          \inst{1}
   }

   \offprints{Martin Haas, haas@astro.rub.de}

   \institute{Astronomisches Institut Ruhr--Universit\"at Bochum,
              Universit\"atsstra{\ss}e 150, 44801 Bochum, Germany
              \and
             Department of Physics, University of California, Santa
             Barbara, CA 93106, USA
             \and
             European Southern Observatory,
             Karl--Schwarzschild--Str. 2, 85748 Garching, Germany
             \and
             Th\"uringer Landessternwarte Tautenburg, Sternwarte 5,
             07778 Tautenburg, Germany
   }

   \date{Received 17. June 2009; accepted 05. Aug. 2009}

  \abstract
  {
    This is the fourth in a series of papers based on the ISOCAM 
    Parallel Survey at 6.7~$\mu$m. While the first three papers have
    been devoted to active galactic nuclei (AGN), here we report on
    emission-line galaxies without AGN signatures in their optical
    spectra.  
  }
  { Polycyclic aromatic hydrocarbon (PAH) emission has been found 
    in both starbursts and modestly starforming galaxies, but the
    relation between starforming activity and PAH luminosity is 
    still a matter of debate. The different correlation degrees
    could be caused by the variety of optical
    and far-infrared sample selection criteria.
    In order to obtain a census of the typical properties of
    PAH emitting galaxies, we here study moderately distant 
    galaxies which have been selected by their PAH emission. 
  }
  {
    Combining the ISOCAM Parallel Survey at 6.7~$\mu$m with 2MASS 
    we have colour-selected a sample of 120~candidates for strong
    PAH emission.
    We obtained optical and 
    mid-infrared  spectra of 75 and 19~sources, respectively,
    and analysed IRAS-ADDSCANs
    and available Spitzer 3.6-160\,$\mu$m photometry.
  }  
  {
    The Spitzer mid-infrared spectra exhibit clear 
    PAH features and corroborate that our photometric selection 
    criteria trace the PAH emission of galaxies fairly well.  
    The optical spectra show emission lines, 
    at median redshift $z$ $\sim$ 0.1, 
    as well as H$_{\delta}$ and Ca\,II absorption, 
    indicating ongoing starformation as well as post-starbursts. 
    The mid- and far-infrared spectral energy distributions (SEDs)
    provide evidence for a broad range of far-infrared (FIR) luminosities
    ($10^{9} L_{\odot} - 2 \cdot 10^{12} L_{\odot}$), 
    but in general 
    the dust is colder ($T \la 25\,K$, $\beta=2$)
    than for starburst galaxies like M\,82 ($T \approx 40\,K$). 
    For most galaxies the monocromatic luminosity $(\nu \cdot L_{\nu})$ peaks
    at about equal height at optical, 6.7\,$\mu$m (PAH) and FIR
    wavelengths. 
    In about 15\% of the sources the FIR luminosity exceeds the
    optical and  PAH energy output by about a factor 5--10 despite the cool 
    dust temperature; in these galaxies a large 
    dust mass of 10$^{8}$ - 10$^{9}$\,$M_{\odot}$ is inferred. 
  } 
  {
    At moderate distance ($z\sim0.1$), PAH selected galaxies 
    turn out to be a quite heterogeneous population of dust-rich, 
    partly infrared-luminous galaxies but mostly cool with a range of 
    post-starburst signatures and starforming activity 
    which appears to be rather modest relative to the entire gas
    content
     (derived from the dust mass and assuming a standard 
      gas/dust ratio). 
      Our results on PAH selected galaxies question the often
      expressed interpretation that the majority of
      high redshift galaxies detected in 15 and 24\,$\mu$m surveys 
      are dominated 
      by powerful ongoing starbursts with high starforming efficiency. 
  } 
  \authorrunning{M. Haas et al.}
  \titlerunning{ ~~PAH selected galaxies}
  \maketitle
%

\section{Introduction}

Polycyclic aromatic hydrocarbonates (PAHs) emit prominent 
features in the mid-infrared around 6-9\,$\mu$m (Puget \& Leger 1989).
PAH carriers are widely distributed in the dusty interstellar medium
of our Galaxy (Mattila et al. 1996) and they are excited in the 
photodissociation regions (D. Cesarsky et al. 1996) as well as by the
mild UV-radiation field around A- and F-type stars 
(Lemke et al. 1998, Uchida et al. 2000). 
PAH emission occurs in normal spirals (e.g. Mattila et al. 1999) 
as well as in luminous and ultra-luminous starbursts (ULIRGs, 
Genzel et al. 1998). PAH features have been detected also in 
dusty elliptical galaxies (Bregman et al. 2008, Kaneda et al. 2008) 
and in AGN (Schweitzer et al. 2006, Shi et al. 2007), 
although faint above the underlying
continuum.

In the local universe ($z<0.01$), the PAH flux 
appears to be better correlated with the 850 and 160\,$\mu$m emission 
from cold dust than with the 15 and 24\,$\mu$m warm dust emission 
(Haas et al. 2002, Bendo et al. 2008). Compared with normal galaxies,  
ULIRGs show an enhanced 100$\mu$m/PAH flux ratio (Klaas et al. 2001),
and among optically selected metal-rich starforming galaxies 
the 70$\mu$m\,/\,8$\mu$m flux ratio increases with luminosity 
(Monkiewicz et al. 2008). 
An analysis of Revised Shapley-Ames galaxies in the SINGS survey suggests 
that it is uncertain by a factor of 10 to use the PAH 
luminosity to extrapolate the far-infrared (FIR, $>$\,60\,$\mu$m) 
properties and the starforming activity (Dale et al. 2005). 
For example, consider two galaxies with the same PAH luminosity: 
A quiescent one like NGC\,891 with a large amount of cool dust 
(SED peaking at 120\,$\mu$m, with star forming efficiency 
$L_{IR}/M_{gas} \sim 1~L_{\odot}/M_{\odot}$, Chini et al. 1986) 
and an active one like M\,82 with about a factor of 10 
less dust mass which is heated to higher temperatures (SED 
peaking at 60$\mu$m, with star forming efficiency $L_{IR}/M_{gas} \sim 50~
L_{\odot}/M_{\odot}$, Thronson et al. 1987). 
Despite their similarity in PAH luminosity, 
the galaxies are clearly different in FIR luminosity and starforming 
activity, hence different in physical states.

Because the underlying continuum from stars and very small dust grains 
is relatively weak (except for AGN and elliptical galaxies), 
the 6-9\,$\mu$m PAH emission of local galaxies dominates by far 
($\ga$\,80\,\%) the 
total flux seen in ISOCAM 6.7\,$\mu$m and Spitzer 8\,$\mu$m broad 
band images.  
At cosmological distances the PAH emission shifts into 
the ISOCAM 15\,$\mu$m ($z$=1) and Spitzer 24\,$\mu$m ($z$=2) 
passbands and could dominate the observed fluxes as well. 
This could naturally explain the peak found in the differential
galaxy number counts, without invoking an extraordinary new galaxy 
population (Xu 2000).  
While for high-redshift galaxy populations (detected in the 
GOODS fields in the 15 and 24\,$\mu$m passbands) 
the rest-frame FIR observations are in general not available, 
models derived from local - mostly IRAS detected - templates 
suggest also a high far-infrared luminosity 
(e.g. Elbaz et al. 2002, Caputi et al. 2007).  
But the key question remains open as to whether PAH emitting galaxies  
typically show an intense M\,82-like starburst activity, presumably 
affecting the whole galaxy, 
or whether the sources are large dust-rich systems with 
moderate star formation relative to the total gas content. 
Therefore, it is vital to establish the nature of PAH emitting 
galaxies over a range of luminosities and distances. 

Combining the spectroscopic SDSS galaxy sample with the Spitzer
First Look and SWIRE surveys,
a good correlation between 8$\mu$m, 24$\mu$m and 70$\mu$m luminosity 
has been reported (Wu et al. 2005, Zhu et al. 2008). However, 
these samples are optically selected to have strong H$_{\alpha}$
emission and, because SWIRE reaches much deeper than SDSS, a substantial 
fraction of PAH emitting IR galaxies may have been excluded; 
for example these samples do not contain any ULIRGs.  
Therefore, it may be put into question as to how far these correlations 
from optically selected samples  
are relevant for properly selected PAH emitting galaxies. 

The ISO-ELAIS survey contains some fields observed at
6.7\,$\mu$m (e.g. V\"ais\"anen et al. 2002), but so far
no specific results on PAH emission have been reported, 
since the survey focused on exploring the cosmic starforming
history to higher redshifts mainly using the 15 and 90\,$\mu$m data 
(Rowan-Robinson et al. 2004 and references therein).   
The ISO-FIRBACK survey at 170\,$\mu$m has found a population 
of IR-luminous but cold galaxies, rather neglected up to now 
(Patris et al. 2003, Dennefeld et al. 2005). 
While 22 of these FIR selected galaxies with clean counterparts at shorter
wavelengths show PAH emission as inferred from Spitzer-IRAC 
photometry, the modelled $L_{3-1000 \mu m}$/$L_{PAH}$ ratio in total spreads 
by about a factor of ten (Sajina et al. 2006).

A proper study of the census of PAH emitting galaxies should be 
built on galaxy samples selected by their PAH emission. 
We here report such a study at moderate distance (z$\sim$0.1) 
based on the ISOCAM 6.7$\mu$m Parallel Survey.
Distances in this paper are calculated using a $\Lambda$CDM cosmology with
$H_0 =71$~km/s/Mpc, $\Omega_{{\rm m}} = 0.27$
and $\Omega_{\Lambda} = 0.73$.

%

\section{Sample selection}

\subsection{ISOCAM 6.7$\mu$m Parallel Survey and 2MASS}
The most prominent PAH emission features lie around 7$\mu$m (with
peaks at 6.2, 7.7 and 8.6$\mu$m) and at 11.3\,$\mu$m.  
The bandpass of the  ISOCAM $LW2$ filter is 5-8.5$\mu$m centered at 6.7$\mu$m.
Thus, $LW2$ is able to measure 6.2\,$\mu$m PAH emission
for moderately distant galaxies up to redshift $z \sim 0.3$. 

Parallel to the observation of a prime target by ISO's photometer
ISOPHOT or the spectrographs ISOSWS/ISOLWS, the mid-infrared 
camera ISOCAM (C. Cesarsky et al. 1996) randomly mapped sky locations  
next to the prime target using the $LW2$ filter.
This results in the 6.7$\mu$m ISOCAM Parallel Survey. It  
covers a total area of 27 deg$^{\rm 2}$
and contains about 16000 point sources (FHWM=5") with a flux limit
down to F(6.7$\mu$m)\,$\sim$\,1\,mJy
(Siebenmorgen et al. 1996, 2000, Ott et al. 2003, 
Ott el al. in preparation).
At high galactic latitude $|b|$\,$>$\,20$^{\circ}$, 
the surveyed area is about 10 deg$^{\rm 2}$.
It provides a suitable hunting ground to search for 
galaxies with prominent PAH emission. 

In order to select such galaxies, we have matched  
( within  a radius of 2$\arcsec$) the ISOCAM sources 
with the 2MASS point source catalog, with the USNO-B, DSS and UCAC optical
catalogues, as well as the NVSS and FIRST radio surveys, and analysed
IRAS ADDSCANs.
The point source criterion ensures that the galaxies are compact
and/or at sufficient distance to be unresolved by these two surveys 
(5" angular extend corresponds to
$\sim$5\,kpc at $z$=0.1).  
We excluded objects which have multiple NIR and
optical counterparts within 10$\arcsec$, or are contaminated by extended
sources (2MASS XSC), or have proper motion ($pm > 3\,\sigma$ from
UCAC). 
At $|b|$\,$>$\,20$^{\circ}$, this yields about 3000 isolated ISO-2MASS 
sources with clean photometry. 
The brightness of the sources goes down to $B \sim 20.5\,$mag,
$K_{\rm s} \sim 15.5\,$mag, $LW2 \sim 12\,$mag (Vega based system).
The typical uncertainties on the $H - K_{\rm s}$ and $K_{\rm s} - LW2$
colours are about 0.2 mag and 0.4 mag, respectively. 
Essentially none of the sources is listed in the NVSS and FIRST radio
catalogues. 
Note that our procedure may have excluded closely interacting galaxy pairs.  

\begin{figure}
  \includegraphics[angle=0,width=\columnwidth,clip=true]{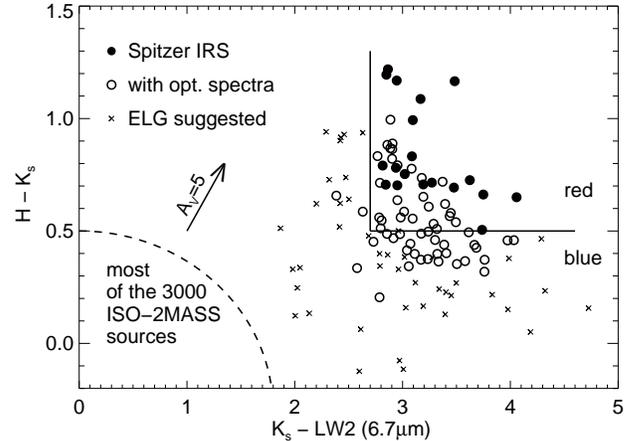}
  \caption{
    Colour-colour diagram illustrating the sample selection.
    While most of the 3000 ISO-2MASS sources lie in the region around
    zero outlined by the dashed curve, 
    here we consider only sources with 6.7\,$\mu$m excess 
    over 2.2\,$\mu$m emission ($K_{\rm s} - LW2 > 1.8$). 
    We obtained optical spectra for all sources in the upper right 
    corner (red subsample, indicated by the solid 
    lines) and a random subset of the remaining sources.
    Sources without optical spectra are suggested emission line
    galaxies (ELGs).
    Spitzer IRS spectra were taken for the brightest red 
    sources (filled circles). 
    AGN would populate the red colour range; they are not shown here. 
    Local starforming galaxies like NGC\,891, NGC\,6946, Arp\,244 and
    M\,82 have $H-K_{\rm s}<0.5$ and $K_{\rm s} - LW2 > 2.7$.
    \label{fig_sample_selection}
  }
\end{figure}

\subsection{Near-mid-infrared colour selection}

The source selection is described in more detail by Haas et al. (2004,
Paper~1). 
We give a brief desciption here: stars and passive
elliptical galaxies populate the colour range $K_{\rm s}-LW2 < 1$.
About 150 of the 3000  ISO-2MASS sources show a 6.7 $\mu$m
flux excess which we defined by $K_{\rm s}-LW2 > 1.8$ 
(Fig.~\ref{fig_sample_selection}).
We consider them 
as candidates for galaxies with strong PAH emission
(hereafter named PAH-candidates).
The threshold $K_{\rm s}-LW2 > 1.8$ is based on the comparison with
local normal galaxy populations (Dale et al. 2001) as well as 
luminous and ultra-luminous infrared galaxies (Genzel et al. 1998, 
Klaas et al. 2001, Laurent et al. 2000) for
which we found suitable $LW2$ data in the ISO Data
Archive\footnote{http://www.iso.vilspa.esa.es/ida/}. 
While among local galaxies only ultra-luminous infrared galaxies 
(ULIRGs) have red $H-K_{\rm s}>0.5$, at
redshift $z\ga 0.1$ the $K$-correction may shift normal galaxies to 
red $H-K_{\rm s}$ 
values.\footnote{http://www.ipac.caltech.edu/2mass/releases/sampler/sampler.html}

Active galactic nuclei (AGN) populate the 
colour range $K_{\rm s}-LW2 > 2.7$ and, at redshift $z<1$, also 
$H-K_{\rm s} > 0.5$ (Haas et al. 2004). 
While this range overlaps with that of the PAH-candidates,
in this paper all sources with AGN signatures in their 
optical spectra are omitted.    
The results on the 33 ISO-2MASS-AGN 
have been presented by Leipski et al. (2005, 2007, Papers~2+3). 

The sample of 120 PAH-candidates is listed in Table \ref{table_sample}.  
None of the 75 sources with optical spectra turned out to be a star. 
Therefore we adopt that the sample 
consists of 44 "red" galaxies with colours similar to those of AGN,
and 76 "blue" galaxies outside the AGN colour range.

\section{Follow-up data}

\subsection{Optical spectroscopy}
In order to study the emission- and absorption-line properties 
of the PAH-candidates and to determine their redshifts,  
we have performed optical spectroscopy in the wavelength 
range between 3500 and 9000 $\AA$.
As mentioned above, among the red subset also AGN are expected. 
Therefore,  
we obtained spectra for the complete set of 44 red sources. 
For the remaining subset of 76 blue sources (120 - 44)
we restricted the observational effort and obtained optical spectra 
of 19 sources randomly selected. 
Among these blue sources, no AGN was found. 

The
data were obtained during 2004 and 2005 at various
telescopes\footnote{South
African Astrophysical Observatory 1.9m, Tautenburg 2m, Kitt Peak 2.1m,
Calar Alto 2.2m, Nordic Optical Telescope 2.5m, Telescopio Nazionale
Galileo 3.5m,
ESO/NTT 3.5m, CTIO Blanco 4m, ESO/VLT 8.2m}, and standard data reduction
was applied. The quality of the spectra is not homogeneous, 
but in general, it is sufficient to
identify the prominent emission lines.
Faint lines or absorption structures are measurable only 
in the best spectra of the brightest objects. 
Examples are shown in Fig.~\ref{fig_optical_spectra}.
 
Good spectra of additional 12 sources are available from the 
Sloan Digital Sky Survey (SDSS, data release 7), as well as  
photometry of 43 sources for which we also inspected the morphology 
(Table~\ref{table_sample}). 

\subsection{Mid-infrared spectroscopy}
\label{sect_data_mir_spectroscopy}

In order to verify the PAH nature of the 6.7$\mu$m emission
and to exclude the possibility of buried  AGN, 
for the 19 brightest red sources we obtained low-resolution
mid-infrared (MIR)
spectra using the Infrared Spectrograph (IRS, Houck et al. 2004)
onboard the Spitzer Space Telescope (Werner et al. 2004).

Spectra of 9 sources were observed in cycle-1 (prog-id 3231) 
covering the full
wavelength range 5-38 $\mu$m with a relatively short integration time
(4$\times$14s in SL and 4$\times$30s in LL).
All 9 spectra show
clear PAH features but poor S/N in the high-excitation emission lines
used for
excluding or establishing buried AGN. 
Therefore, in cycle-2 (prog-id 20090) 
we observed 10 more sources with longer 
integration time (12$\times$120s), but only in the range 19-38
$\mu$m containing the high-excitation lines [NeV]$_{\rm 24.3 \mu m}$ and 
[OIV]$_{\rm 25.9 \mu m}$. 
All 19 spectra were taken without peak-up images.

Starting from the background subtracted 
pipeline products, we performed standard interactive data
reduction, using the SPICE software tool as well as our 
own routines for improved
cleaning of residual rogue pixels and and cosmic rays, 
and for combining the spectra of
the four IRS channels. 

Table~\ref{table_mir_continuum} lists the MIR spectral properties
as well as continuum fluxes at 15, 24 and 35\,$\mu$m. 

Integration of the spectra over the $LW2$ passband yields a good
coincidence between the $LW2$ and IRS photometry for all 
but one of the 9 sources 
($F_{\rm LW2} / F_{\rm IRS} = 1.00 \pm 0.14$).\footnote{ 
The outlier (2MASS\,15554606$+$1532218) has a factor 1.8 lower IRS flux in 
the $LW2$ passband. The discrepancy could be explained by 
either a photometric $LW2$ error due to a non-detected cosmic ray event,
or an IRS pointing error or aperture effects.
On SDSS images the source is extended with a diameter of
about 8" including a bright tail to the south. 
While from the $LW2$ image the total flux is measured, 
part of the flux could be missed in the IRS spectrum due to the 
slit width of 3.7" between 5\,$\mu$m and 14\,$\mu$m.
However, the IRS spectrum of 2MASS\,1555 does not show a jump 
at 14\,$\mu$m longward of which the slit width is 11.6$\arcsec$.
Probably a combination of all three effects is responsible for the 
discrepancy between $LW2$ and IRS photometry of this source. 
Because this source has typical average properties, 
the results and conclusions are essentially not affected by 
the choice of the 6.7\,$\mu$m flux value. 
}

\subsection{Mid- and far-infrared photometry}
\label{sect_mfir_photometry}

All but 4 sources are not contained in the IRAS faint
source catalogue. Therefore we analysed IRAS
ADDSCANs\footnote{http://scanpi.ipac.caltech.edu:9000/}.
Using a matching radius of 20$\arcsec$ around the 2MASS position, 
19 sources could be detected with at least 4$\sigma$ 
at 60\,$\mu$m and 13 of them also at 100\,$\mu$m. 
The fact that 13/19 sources are detected in both filters supports that the
detections are real. 
Because the flux level is rather low, we adopt in general a 
30\% photometric uncertainty, although
for 4 sources listed in the IRAS faint source catalog 
our ADDSCAN-photometry agrees within 1\%. 
We also checked on the 2MASS and ISOCAM images 
that no nearby ($<$30$\arcsec$) red source is likely to contribute 
to the IRAS fluxes. 
(Actually two more sources had been detected on ADDSCANs, 
but they have been discarded.)
Table~\ref{table_sample} lists the photometry 
of the accepted detections 
and the 60\,$\mu$m upper limits of the remaining sources. 

In the wavelength range 3.6-160\,$\mu$m, 
14 sources are covered
on archival maps taken with the Spitzer Space Telescope. 
All sources are detected in all bands observed 
and match well with the 2MASS positions 
(better than 2$\arcsec$ for IRAC 3.6-8\,$\mu$m and 
4-10$\arcsec$ for MIPS 24-160\,$\mu$m, respectively).  
Using the Post Basic Calibrated Data 
Products, we derived aperture photometry with an uncertainty 
of about 10\% (IRAC) and less than 30\% 
(MIPS). 
Because the used apertures are sufficiently large, e.g. 12$\arcsec$
  for IRAC, no aperture corrections are necessary.
The values are listed in Table~\ref{table_spitzer_photometry}. 

We computed far-infrared luminosities using the standard formula 
(Sanders \& Mirabel, 1996, their Table~1): \\
$L_{FIR}$ = 
4$\pi$~$D_{L}$$^{\rm 2}$ $\times$ 1.6 $\times$ 1.26 $\times$ 10$^{\rm -14}$ (2.58 $F_{60}$ + $F_{100}$)  [$L_{\odot}$],
with $F_{60}$ and $F_{100}$ given in Jy. 
For the 6 cases with 60\,$\mu$m detection but 100\,$\mu$m upper
limit, we adopted $F_{100} = 2 \times F_{60}$. 
$F_{100}$ is probably even higher, because most sources have 
$F_{60}/F_{100}>2$, if measured. 
Note that the $L_{FIR}$ formula assumes an average star forming galaxy SED 
and thus may weaken possible differences between M\,82- and NGC\,891-types. 
But given the photometric uncertainties,
this approximation should be sufficient for our purpose, and  
any difference between the IRAS 60+100\,$\mu$m and Spitzer 
70+160\,$\mu$m passbands should play a minor role. Therefore, 
we used the same formula also for the Spitzer 70 and 160\,$\mu$m data to
derive $L_{FIR}$, adopting $F_{60} \approx F_{70}$ and $F_{100} \approx F_{160}$
and  $F_{160} = 2 \times F_{70}$, when $F_{160}$ is 
not observed. 

\subsection{K-correction}
\label{sect_k_corrections}
Because our sources have strong PAH features as shown below, 
$K$-correction plays a role. 
\begin{itemize}
\item[1)]
We have determined the $K$-correction for $LW2$ fluxes  
from local ISO-SWS spectra of starforming galaxies. 
For the redshift of our sources  from $z=0.03$  to $z =0.3$ 
the $K$-correction factor increases steadily from 1.1 to 2.2. 
\item[2)]
For 24\,$\mu$m fluxes we determined $K$-correction factors 
from our 19-38\,$\mu$m spectra. 
To correct also the photometry data of sources without spectra, 
a fit yields\\ 
F24 ($K$-corrected) = $(1+3z) \times$ F24 (observed). 
\item[3)]
In the FIR $K$-correction plays a minor role, at least with regard to the
measurement errors. 
At 60 and 70\,$\mu$m we applied a  $K$-correction factor of
{\bf 
$(1+0.33z)$, i.e. 0 at $z=0$ and 10\% at $z=0.3$, 
} 
and no $K$-correction at 100 and 160\,$\mu$m. 
\end{itemize}
All data points shown in Figures \ref{fig_col_mag} to \ref{fig_luminosities} 
have been $K$-corrected.


\section{Results and discussion}

\begin{figure}
  \includegraphics[angle=0,width=\columnwidth,clip=true]{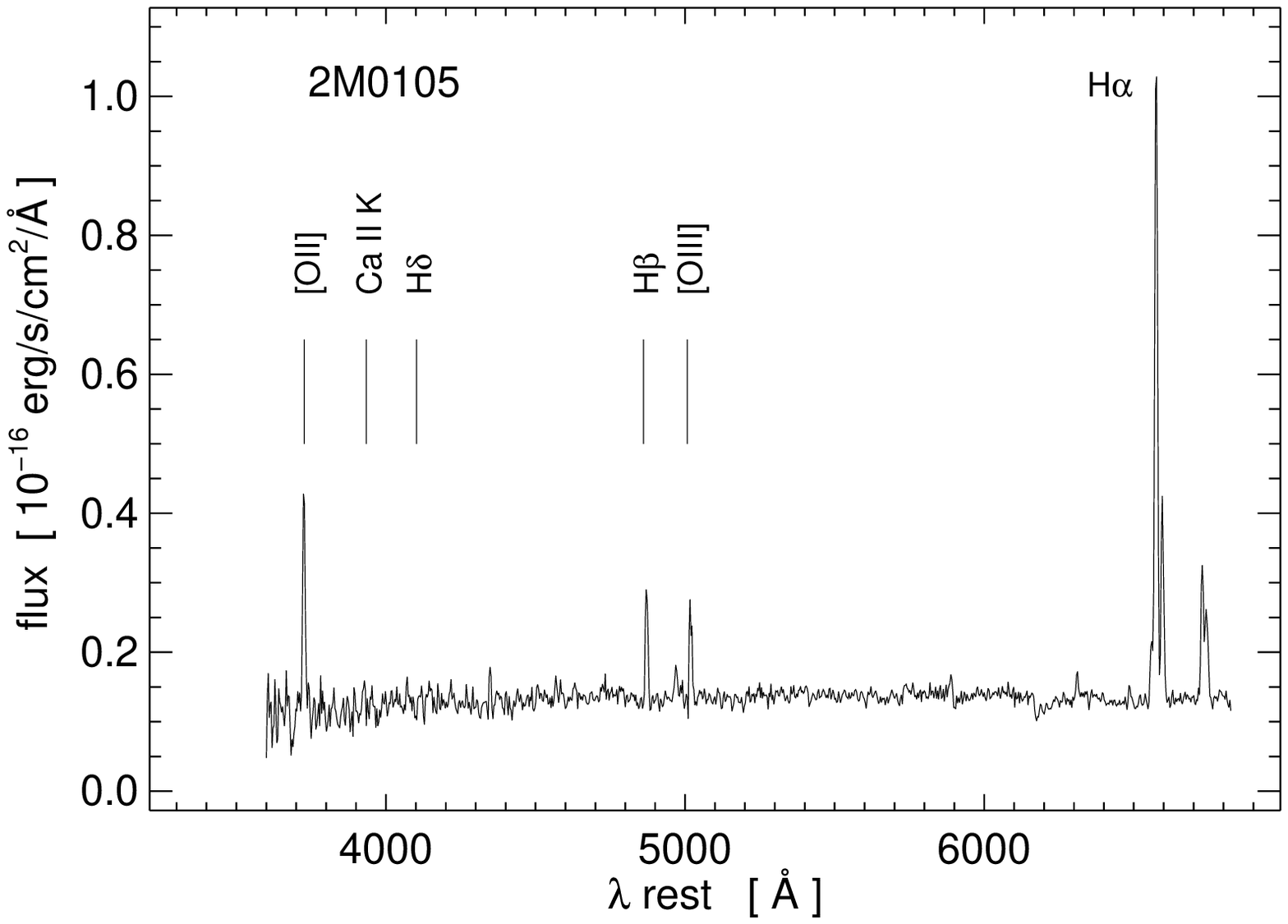}
  \includegraphics[angle=0,width=\columnwidth,clip=true]{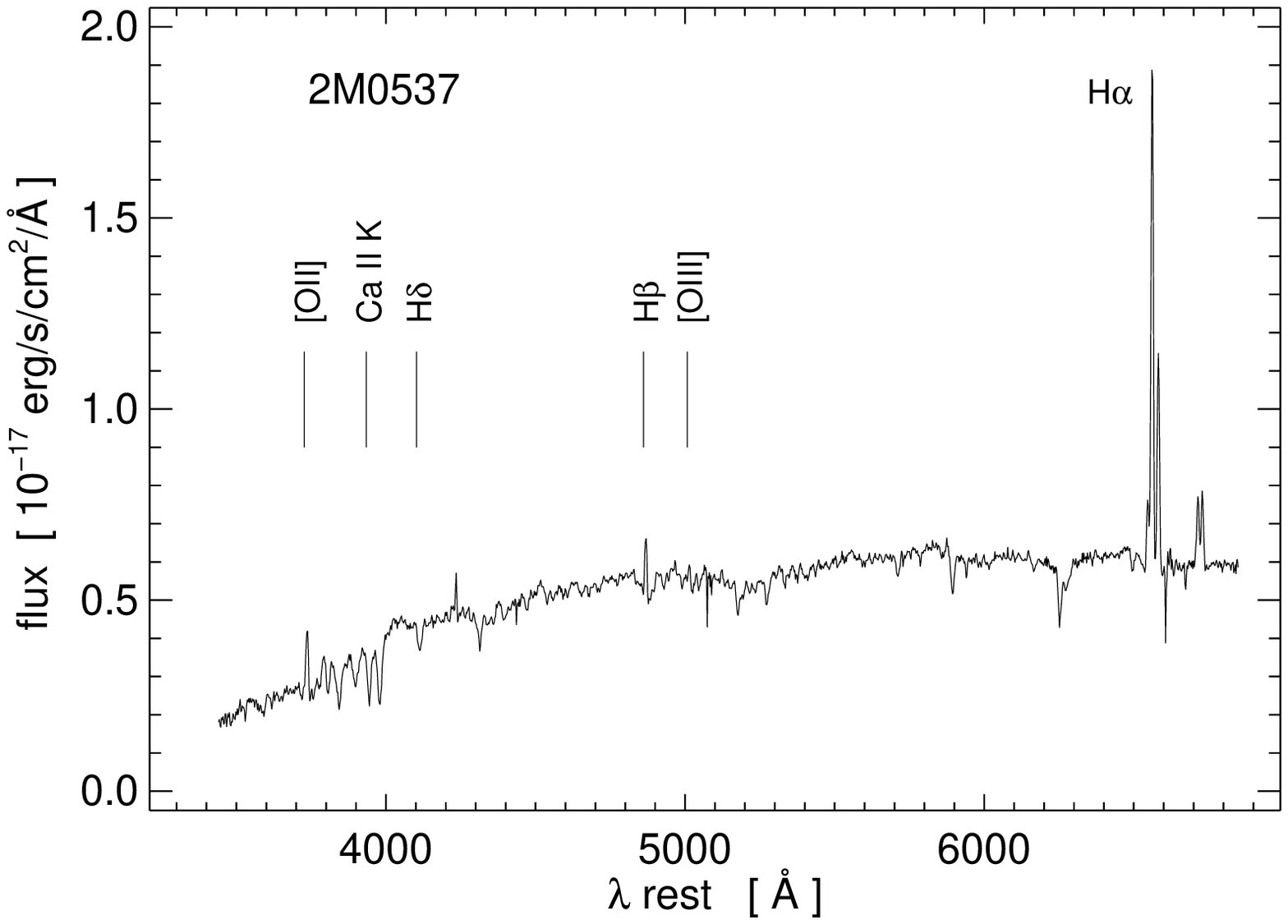}
  \caption{
   Examples of optical spectra. 
   Top (2MASS 01051501 $-$2612466):
   While H$_{\alpha}$ and [O\,II] are prominent, 
   H$_{\beta}$ appears suppressed. 
   Bottom (2MASS 05375494 $-$4408547): 
   While H$_{\alpha}$ is well seen, the other emission lines
   appear relatively faint and the H$_{\delta}$ and
   Ca\,II\,K absorption is rather strong.    
    \label{fig_optical_spectra}
  }
\end{figure}

\begin{figure}
  \includegraphics[angle=0,width=\columnwidth,clip=true]{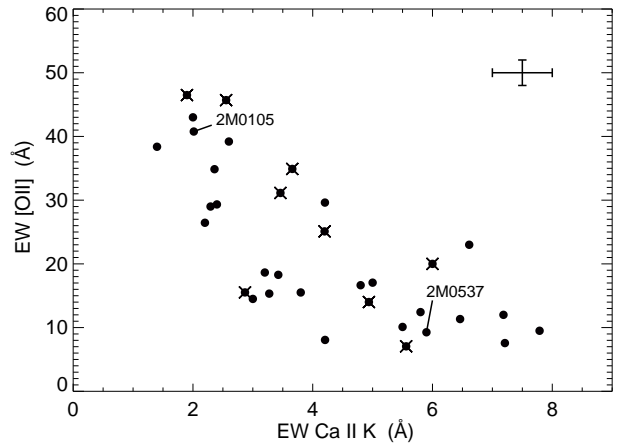}
  \caption{
    Equivalent widths of [O\,II] emission and Ca\,II\,K absorption.
    The average errors are illustrated by the bars in the upper right
    corner. 
    The two sources with spectra shown in 
    Fig.~\ref{fig_optical_spectra} are labeled. 
    Overplotted crosses mark those sources detected with IRAS at
    60\,$\mu$m.
    The  two  sources with the highest EW [O\,II] are the two ULIRGs of our sample. 
    \label{fig_OII_CaII}
  }
\end{figure}

\subsection{Optical spectra}

Two example spectra are displayed in Fig.~\ref{fig_optical_spectra}.
In general, the spectra 
show a range of blue and red continua as well as reddened Balmer
emission lines with a ratio reaching H$_{\alpha}$ / H$_{\beta}$ $>$ 6 even after
correction for stellar absorption troughs. 
Most spectra also show [O\,II]$_{\rm ~3727\AA}$ emission (henceforth simply denoted [O\,II]). 
The equivalent widths of [O\,II] and H$_{\alpha}$ are broadly correlated. 
High-excitation emission
lines ([O\,III]$_{\rm ~5007\AA}$) mostly are not detected or {\bf are} faint, 
and the emission
line ratios place the objects
in the starforming part of the standard diagnostic diagrams used for
separating AGN, LINER and starforming galaxies  
(Veilleux \& Osterbrock 1987).

The equivalent width of [O\,II] is an indicator of the average massive 
star formation rate (e.g. Dressler \& Gunn 1982; Kennicutt 1992). Sodr\'e \&
Stasi\'nska (1999) report a clear correlation of $EW$ [O\,II] with spectral types 
ST obtained with a principal component analysis of the continuum and 
absorption features of spectra. For ST$>$0 their equivalent widths cover
approximately the same range as in Fig.~\ref{fig_OII_CaII} of the present paper with 
$EW \sim 5~\AA$ for ST~0 and $EW \sim 40~\AA$ for 
ST~10.\footnote{The Sodr\'e \& Stasi\'nska (1999) spectral type ST is 
correlated with the Hubble morphological type. ST$>0$ roughly corresponds 
to Hubble types later than Sa, and ST~10 to irregular galaxies.} 
These authors argue that the 
spectral sequence can be interpreted as a sequence of an increasing ratio of 
the present to the average past star formation rate. In that picture, low 
values  of $EW$ [O\,II] correspond, on average, to galaxies where most of the 
star formation occured long ago, very few OB stars contribute to the 
optical spectrum, and the ionizing UV radiation field is dominated by AGB 
stars. The highest values of $EW$ [O\,II] in our sample are measured for the 
two ULIRGS (Sect.~\ref{section_fir}). 

From the 45 $EW$ [O\,II] measurements listed in Table~\ref{table_sample}, 
one finds a remarkably 
high mean value of 23~$\AA$. This can be compared with the mean values
of 11~$\AA$ and 
17~$\AA$ found by Liu and Kennicutt (1995) for a complete, magnitude-limited 
sample of (optically selected) local galaxies and for a sample of more 
distant merger galaxies, respectively. Even for the (certainly unrealistic) 
case that the $\sim$30 galaxies in our spectroscopic sample without $EW$ [O\,II] 
measurements have no [O\,II] emission at all, the sample average of 
14~$\AA$ is larger than that of Liu \& Kennicutt's local sample.

About one third of the objects show H$_{\delta}$ and/or Ca\,II\,K
absorption with equivalent widths $EW > 4~\AA$. 
This indicates that stars of intermediate age contribute strongly 
to the spectra. 
However, there is no galaxy in our 
spectroscopic sample with really strong Balmer absorption indicating a  
dominant A-star population as measured in "E+A" galaxies  with 
$EW H_\delta > 6~\AA$  (Liu and Kennicutt 1995).
The H$_{\delta}$ and Ca\,II\,K absorptions 
suggest that about 0.5--1 Gyr ago an intense starforming episode 
stopped and that we now see the galaxy in a post-starburst
phase, but with considerable ongoing star formation as traced by the 
H$_{\alpha}$ and [O\,II] emission.

The spectra of 34 sources had sufficient S/N at the blue end 
to allow us to quantify both  [O\,II] emission and Ca\,II\,K absorption. 
The equivalent widths of [O\,II] emission and Ca\,II\,K absorption are 
roughly anti-correlated.   
The sources with strong [O\,II] have weak Ca\,II\,K, and vice versa, 
as shown in Fig.~\ref{fig_OII_CaII}. 

So far, we have assumed that 
both the emission lines and their underlying continuum 
suffer, on average, from similar extinction. 
On the other hand, Calzetti (2001) found 
in a UV-selected sample of star forming galaxies that 
the H\,II emission lines undergo a factor $\sim$2 higher 
extinction than the stellar continuum. 
An explanation for this difference could be that H\,II regions 
are located closer to the parent dusty molecular clouds and 
therefore are more likely affected by extinction 
than the bulk of stars. 
%
If the difference between emission line and continuum extinction 
also holds for our sample, then one expects that $EW$ [O\,II] decreases 
with increasing H$_{\alpha}$/H$_{\beta}$ ratio
(which traces the emission line extinction). 
In fact, the mean $EW$ [O\,II] value decreases from $EW$\,=\,26.6\,\AA~  
for sources with low H$_{\alpha}$/H$_{\beta}$ $<$ 6 to $EW$\,=\,14.2\,\AA~ 
for high H$_{\alpha}$/H$_{\beta}$ $>$ 10. 
However, the population in the lower right 
corner of  Fig.~\ref{fig_OII_CaII}
($EW$ [O\,II] $<$ 20 and $EW$ Ca\,II\,K $>$ 4) contains sources with 
low as well as high emission line extinction which again is not
correlated with IRAS 60\,$\mu$m detection or non-detection. 
In addition, the highest $EW$ [O\,II] are seen in the two ULIRGs
although they have H$_{\alpha}$/H$_{\beta}$ $\ge$ 10.
This indicates that, even in the case of 
high emission line extinction, the [O\,II] equivalent width can identify
the star forming activity fairly well and consistent with MIR emission
line diagnostics (Sect. \ref{sect_mir_sf_tracers}).  
Furthermore, 
the $EW$ [O\,II] comparison of our sample with that of others 
(Sodr\'e \& Stasi\'nska 1999, Kennicutt 1992, Liu and Kennicutt 1995)
was performed in a consistent manner, all samples not being 
extinction corrected. 
These considerations give us confidence  
that the potential extinction difference 
between emission lines and continuum  
does not affect our basic conclusions on the wide range of starforming 
activity and the presence of aged stellar populations in PAH selected
galaxies.  


We inspected the extent and morphology of 43 sources with 
SDSS images available.
Most (34/43) of the sources appear extended ($>5\arcsec$) and about one
third (12/43) shows irregular morphology or 
possible faint companions (with unknown redshift). However the 
data did not allow us to establish a connection 
between morphology and spectral properties such as equivalent widths.
Nevertheless, this suggests that disturbed morphology does not
automatically imply the presence of powerful starbursts. 
  
To summarise, the optical spectra indicate a range of ongoing 
starformation as well as evidence for a intermediate-age stellar populations. 
The basic conclusions have been drawn 
without calculating starforming rates from the emission line 
fluxes, because the corrections for extinction and for aperture 
loss due to the $1-2\arcsec$ slit widths introduce large uncertainties. 

\subsection{Mid-infrared spectra}
\label{sect_mir_spectra}

\subsubsection{PAH nature of the 6.7\,$\mu$m flux}
\label{sect_pah_nature}

While local non-AGN templates suggest that PAH emission 
dominates the  6.7\,$\mu$m flux of our sources, 
we need to exclude that hidden AGN contribute 
significantly to the 6.7\,$\mu$m continuum. 
Hard X-ray data ($> 2keV$) are not available for our sample. 
Therefore, we checked the nature of the 6.7\,$\mu$m flux 
by means of MIR spectra. 

\begin{figure}
  \includegraphics[angle=0,width=\columnwidth,clip=true]{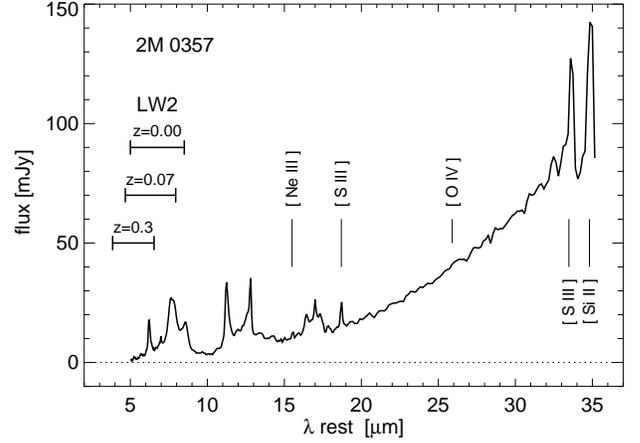}
  \caption{
    Example IRS spectrum of 2MASS\,03574895$-$1340458 exhibiting strong PAH
    features at 6-9~$\mu$m and also at 11.3, 12.7 and 17~$\mu$m.
    Three emission lines are clearly detected: 
    [S\,III] at 18.71 and 33.48~$\mu$m and [Si\,II] at 34.81~$\mu$m,
    while [Ne\,III] at 15.55~$\mu$m is very faint and [O\,IV] at 25.89~$\mu$m is not detected.
    The horizontal bars indicate for three redshifts how much of the
    PAH features is covered by the $LW2$ passband.
    \label{fig_irs_of_isocp}
  }
\end{figure}

Figure~\ref{fig_irs_of_isocp} displays the IRS spectrum of the
brightest source 2MASS\,03574895$-$1340458. The
5-38\,$\mu$m spectra of the other 8 cycle-1 sources  
look very similar. 
The MIR spectra show strong 6-9\,$\mu$m PAH emission. 
Depending on the redshift of our sources ($0.03<z<0.3$), 
the $LW2$ passband essentially catches
the 6.2\,$\mu$m PAH feature and part of the 7.7\,$\mu$m feature. 
The 8.6\,$\mu$m PAH feature is shifted out of $LW2$. 

On average, about 50\% of the rest frame 6-9\,$\mu$m PAH flux 
falls into 
the $LW2$ passband and thus explains why the objects have high 
$K-LW2>1.8$ colour values and meet the selection criterion. 
The individual values of the PAH fraction seen in $LW2$ are 
listed in Table\,\ref{table_mir_continuum}.
Among our sample, 
however, we do not see a trend of $K-LW2$ colour with redshift,
indicating that the broad range of $K-LW2$ (Fig.~\ref{fig_sample_selection})
is not dominated by $K$-correction effects. 

\begin{figure}
  \includegraphics[angle=0,width=\columnwidth,clip=true]{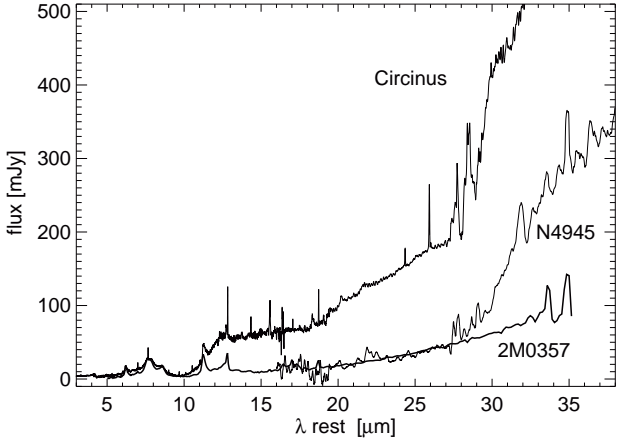}
  \caption{
    IRS spectrum of 2MASS\,03574895$-$1340458 and ISO-SWS spectra
    of the starburst-AGN composite
    Circinus and the elusive X-ray AGN NGC\,4945. 
    The ISO-SWS spectra are scaled to match the PAH flux level at 6-9 $\mu$m. 
    At wavelengths longer than 12~$\mu$m and 30~$\mu$m, respectively,
    the continuum of Circinus and NGC\,4945 is much brighter 
    than that of 2M\,0357. 
    \label{fig_mir_spectrum_with_agn}
  }
\end{figure}

\begin{figure}
  \includegraphics[angle=0,width=\columnwidth,clip=true]{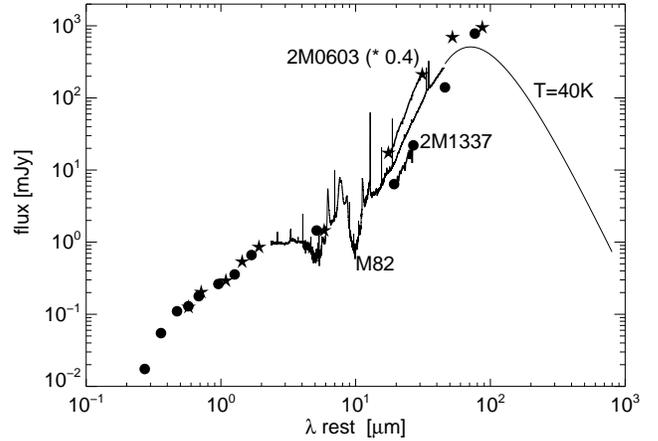}
  \caption{
    SEDs of the two ULIRGs. 
    2MASS\,06033357$-$4509412 (asterisks) is scaled by a factor 0.4 to match 
    the 6.7\,$\mu$m flux of 2MASS\,13371721$+$0904430 (filled circles). 
    In comparison with the spectrum of M\,82, both SEDs show a 
    similarly steep 15-30\,$\mu$m rise.
    The temperature $T=40K$ of the modified greybody is  
    typical for ULIRGs when using an emissivity 
    index $\beta=2$ (Klaas et al. 2001).  
    The optical data are from SDSS.   
    \label{fig_ulirg_sed}
  }
\end{figure}

Any 10\,$\mu$m silicate absorption is difficult to determine,
but if present then it appears weak.
With respect to buried AGN, which in principle could 
contribute to the 6.7\,$\mu$m continuum, we
consider what can be concluded further from the 
19-38\,$\mu$m spectra of cycle-1 and cycle-2.  
High-excitation lines ([NeV]$_{\rm 24.3 \mu m}$ and 
[OIV]$_{\rm 25.9 \mu m}$) are not
detected, even in the ten 19-38\,$\mu$m
spectra with longer integration times. 
The low upper limits of the line luminosities and equivalent widths
argue against powerful buried AGN and against a significant
6.7\,$\mu$m AGN continuum. This conclusion is further
supported when looking for possible re-emission of shielding dust at
longer optically thin wavelengths: With the exception of two ULIRGs, 
the SEDs show only a moderate rise from 15 to 35 $\mu$m, and hence
are different from those of elusive X-ray detected AGN like NGC4945
and of starburst-AGN composites like Circinus 
(Fig.~\ref{fig_mir_spectrum_with_agn}). 
The two ULIRGs of our sample do not have a 5-10\,$\mu$m spectrum, but
cool $F_{25}/F_{60}$ colours (Fig.\,\ref{fig_ulirg_sed}) which argue in
favour of starburst dominated 6.7\,$\mu$m emission with strong PAH
features (Genzel et al. 1998).

These results from 19 sources with MIR spectra lead us to conclude 
that most if not all 
galaxies of our sample show strong PAH emission and that our LW2
photometry serves as a measure for about 50\% of the 6-9\,$\mu$m PAH
luminosity. 

\begin{figure}
  \includegraphics[angle=0,width=\columnwidth,clip=true]{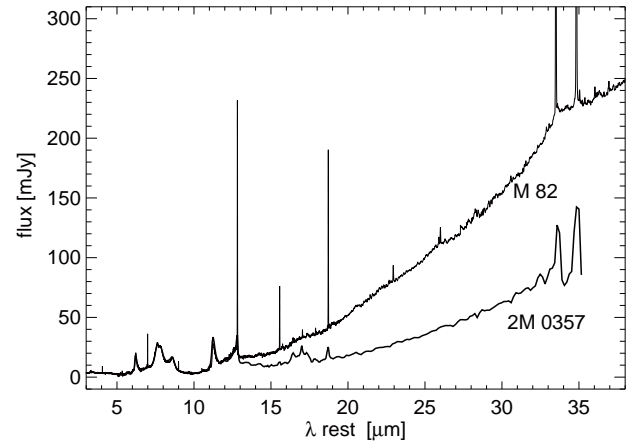}
  \caption{
    IRS spectrum of 2MASS\,03574895$-$1340458 and ISO-SWS spectrum 
    of the starburst template M82. 
    The spectrum of M82 is scaled to match the PAH flux level at 6-9 $\mu$m. 
    At wavelengths longer than 15 $\mu$m 
    the continuum of M82 is much brighter 
    than that of 2M\,0357. 
    \label{fig_mir_spectrum_with_sb}
  }
\end{figure}

\begin{figure}
  \includegraphics[angle=0,width=\columnwidth,clip=true]{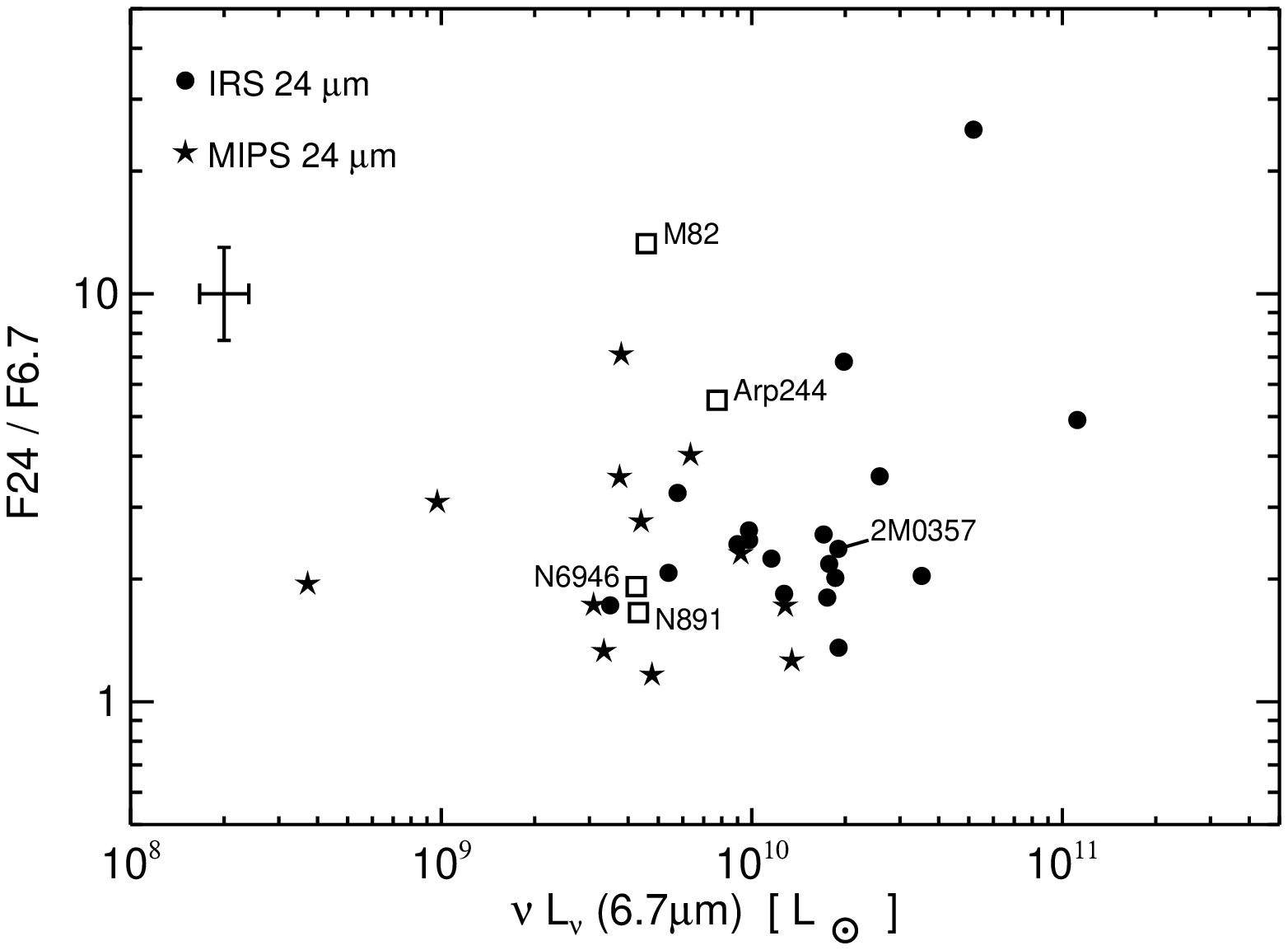}
  \caption{
    Colour-magnitude diagram F24$\mu$m\,/\,F6.7$\mu$m versus 
    6.7$\mu$m luminosity. The data have been $K$-corrected. 
    Uncertainties of 20\% in x and 40\% in y are illustrated 
    in the upper left corner.
    For comparison two spirals (NGC\,891, NGC\,6946),
    the Antennae Galaxy Arp\,244 and the starburst galaxy M\,82 are shown
    (open squares). 
    Also labelled is 2MASS~0357, the source with MIR spectrum 
    shown in Fig.\,\ref{fig_irs_of_isocp}.
    \label{fig_col_mag}
  }
\end{figure}

\subsubsection{Mid-infrared  starforming tracers}
\label{sect_mir_sf_tracers}

[Ne\,II]$_{\rm 12.8 \mu m}$ would be the first choice 
among MIR starforming tracers, 
because of its brightness and the required low excitation potential of
21.6\,eV.
But in the low resolution spectra
this line is not well separated from the 12.7$\mu$m 
PAH feature. Therefore we analysed the 
[S\,III]$_{\rm 18.7 \mu m}$ and [Ne\,III]$_{\rm 15.5 \mu m}$ lines
requiring excitations of 23.3\,eV and 40.9\,eV, respectively.
The fluxes are listed in Table~\ref{table_mir_continuum}.
The [S\,III]\,/\,[Ne\,III] flux ratio lies in the range 2-5 
typical for (modestly) starforming galaxies; for comparison, 
intense starburst galaxies like 
M\,82 have [S\,III]\,/\,[Ne\,III] $\sim$ 2 (Verma et al. 2003)
and the starburst knots in the overlapping region of the Antennae galaxy
pair have [S\,III]\,/\,[Ne\,III] $\sim$ 1 (Kunze et al. 1996).  
This is because the radiation field of bursts of O-stars is harder and a
higher fraction of neon is in the double-ionized state.
There is a marginal trend that sources with high [Ne\,III] fraction also 
show a high equivalent width of H$_{\alpha}$ and [O\,II]$_{\rm ~3727\AA}$. 

Two sources, which turn out to be ULIRGs (Sect.~\ref{section_fir}), 
have steeply rising 20-35\,$\mu$m continua (Fig.~\ref{fig_ulirg_sed}). 
But the MIR continua of all 17 other sources 
do not show the strong rise between 
15 and 35\,$\mu$m typical for warm starburst heated dust like that 
found in M\,82 (Fig.~\ref{fig_mir_spectrum_with_sb}). 

\begin{figure}
  \includegraphics[angle=0,width=\columnwidth,clip=true]{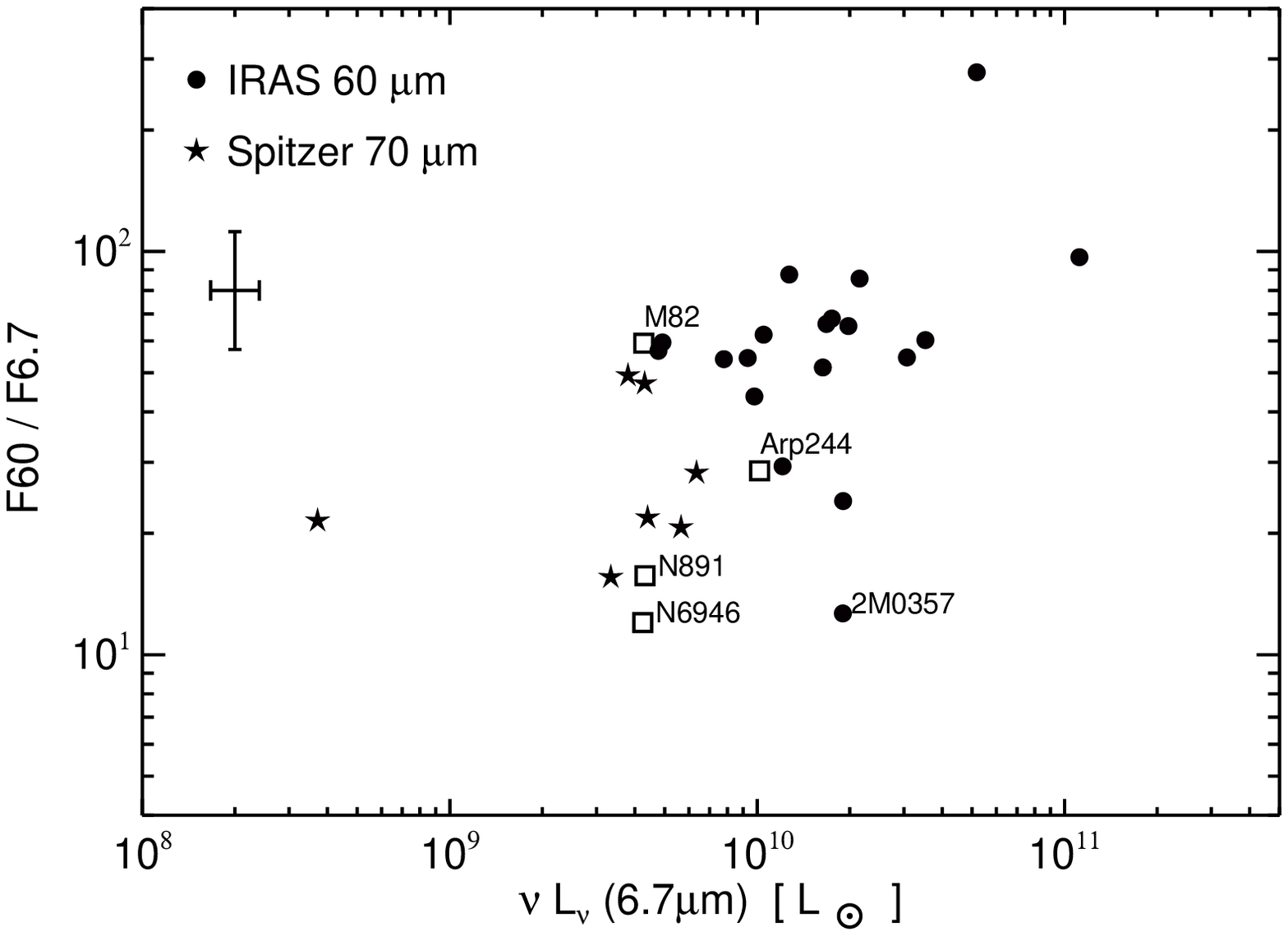} 
  \includegraphics[angle=0,width=\columnwidth,clip=true]{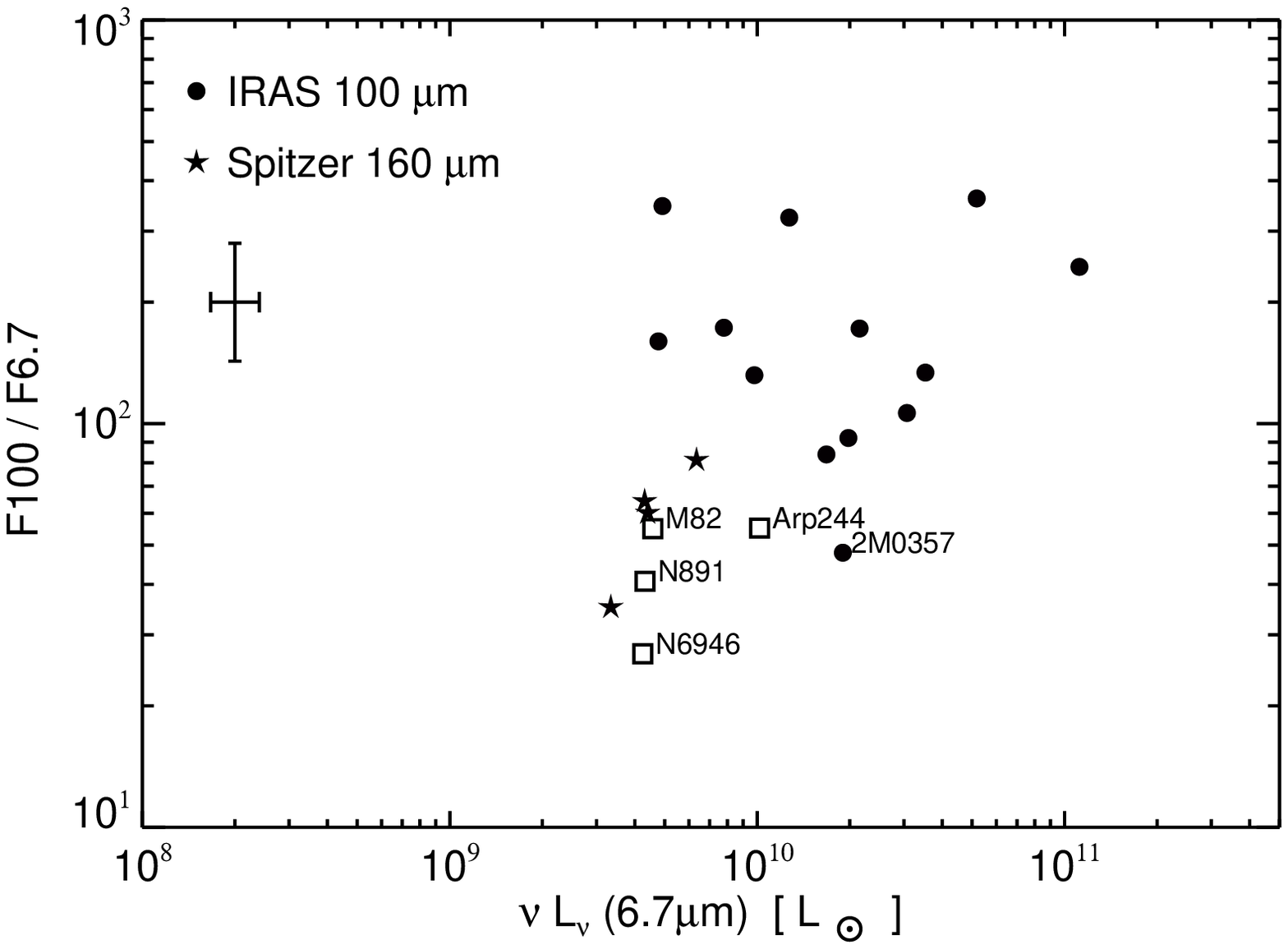}
  \caption{
    Colour-magnitude diagrams F60\,/\,F6.7  
    and F100\,/\,F6.7 versus 
    6.7$\mu$m luminosity. The data have been $K$-corrected. 
    Uncertainties of 20\% in x and 40\% in y 
    are illustrated in the upper left corner.
    For comparison with local templates two spirals (NGC\,891, NGC\,6946),
    the Antennae Galaxy Arp\,244 and the starburst galaxy M\,82 are shown
    (open squares).
    Also labelled is 2MASS~0357, the source with the MIR spectrum shown 
    in Fig.\,\ref{fig_irs_of_isocp}.
    The numerous 60\,$\mu$m upper limits populate the F60\,/\,F6.7
    range between 10 and 100; they are not plotted to avoid confusion.
    \label{fig_col_mag_fir}
  }
\end{figure}

This is also reflected in the low $F_{24}/F_{6.7}$ ratio for most sources
as shown in Fig.~\ref{fig_col_mag}. 
In this plot we included both the $F_{24}$ data from the spectra and the 
MIPS-24\,$\mu$m photometry (Tables~\ref{table_mir_continuum} 
and \ref{table_spitzer_photometry}); 
for three sources without redshift but with photometry  
we adopted the median redshift of the sample ($z=0.105$).
For the comparison sources we used the IRAS 25\,$\mu$m fluxes from the
NED (the difference between IRAS-25 and MIPS-24 bandpasses is
negligible here). 
The $F_{24}/F_{6.7}$  colours of most galaxies appear more similar to 
those of the rather quiescent spiral galaxies 
NGC\,891 (edge-on) and NGC\,6946 (face-on) than to M\,82.   
A few sources reach the $F_{24}/F_{6.7}$ ratio of Arp244, 
the colliding Antennae galaxy pair with 
dust-enshrouded starbursts as well as a large amount of cold dust
(Mirabel et al. 1998, Haas et al. 2000). 

To summarise, 
most ($\sim$15/19) of the mid-IR spectra indicate rather modest starforming activity and 
do not show the signatures of powerful M\,82-like starburst galaxies, 
either in emission lines or the continua.

\subsection{Far-infrared properties}
\label{section_fir}


Figure~\ref{fig_col_mag_fir} shows the 60 and 100\,$\mu$m fluxes
normalized by F6.7\,$\mu$m. 
We include also the Spitzer 70 and 160\,$\mu$m  photometry, 
adopting for the first look that colour corrections play a minor role.      
A striking fact is that at least half of the FIR detected sources show a much 
higher $F_{60}/F_{6.7}$ ratio than the local cool templates,
and the $F_{100}/F_{6.7}$ ratio of virtually all sources exceeds that of
local templates.
Thus the basic result is: 
Apart from the two ULIRGs, 
most if not all of the 17 sources  with IRAS FIR detections are exceptionally 
strong cool dust emitters, relative to their PAH emission.

\begin{figure}
  \includegraphics[angle=0,width=\columnwidth,clip=true]{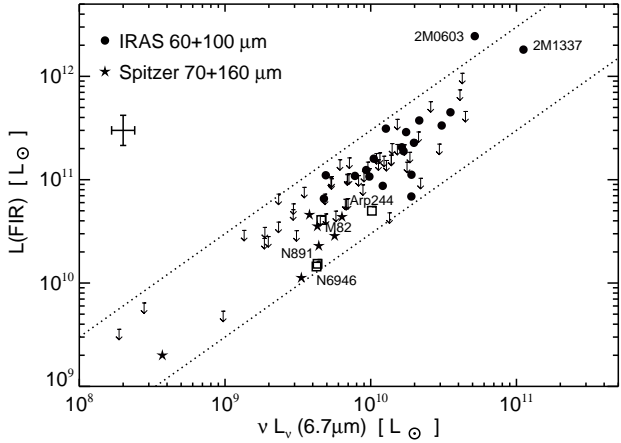}
  \caption{Far-infrared versus 6.7\,$\mu$m luminosities. 
    The data have been $K$-corrected. 
    Upper limits from IRAS  are marked with arrows. 
    Uncertainties of 20\% in x and 40\% in y 
    are illustrated in the upper left corner.
    For comparison with local templates, two spirals (NGC\,891, NGC\,6946),
    the Antennae Galaxy Arp\,244 and the starburst galaxy M\,82 are shown
    (open squares).
    The dotted lines mark the range $L_{FIR}$ = 3 and 30~$\times$~$L_{6.7}$. 
    At highest luminosity, the $L_{FIR}$ / $L_{6.7}$
    ratio tends to increase as indicated by the two labelled ULIRGs. 
    \label{fig_luminosities}
  }
\end{figure}

Further evidence for this conclusion comes from the far-infrared 
luminosities as shown in Fig.~\ref{fig_luminosities}.
Among the 19 IRAS detections 16 sources have $L_{FIR} > 10^{11}
L_{\odot}$. 
Two of them qualify as ULIRGs with $L_{FIR} > 10^{\rm 12} L_{\odot}$
(in Table~\ref{table_sample} they are marked after the 2MASS 
name).\footnote{ 
We do not expect that our sample contains more than these two
ULIRGs. The reason is that 
known local ULIRGs have $K_{s}-LW2>2.7$ and $H-K_{s}>0.5$, and  
all objects of this red subsample have redshifts, so that 
the FIR luminosity can be determined. 
With exception of the two ULIRGs, all other sources of the red
subsample have $L_{FIR} < 10^{\rm 12} L_{\odot}$  even in case of upper
FIR flux limits. Outside of the red subsample a ULIRG could be
expected if $z>0.3$. Then it still has $H-K_{s}>0.5$, 
but $K_{s}-LW2$ becomes lower, because the PAH features move out of the
$LW2$ passband. In this colour range there are 10 sources without
spectra, but for none of them is the optical to MIR SED consistent with a 
ULIRG template at (photometric) redshift $z>0.2$.}  
One of the two ULIRGs (2MASS13371721$+$0904430 at $z=0.3$) has a remarkably high
F100/F60 (=5.5) indicating a large amount of cool dust despite the
ultra-luminous IR emission. A similar cool ULIRG has been found at
$z=0.45$ by Chapman et al. (2002) in the ISO-FIRBACK survey. 

Most of the sources with upper IRAS FIR flux limits 
have $L_{FIR}$ ($< 10^{11} L_{\odot}$). 
The seven Spitzer FIR detections reach about a factor of 5 lower
fluxes than the IRAS upper limits, resulting in both 
lower $L_{FIR}$ and  $L_{FIR} / L_{6.7}$, too,
which makes the sources  similar to local cool templates. 
Therefore, we suggest that most of the sources with IRAS FIR upper
limits also have $L_{FIR} / L_{6.7}$ similar to the Spitzer FIR
detected sources and to local cool templates.  

From Fig.~\ref{fig_luminosities} it is clear that $L_{FIR}$ is
basically correlated with $L_{6.7}$. 
However, the $L_{FIR} / L_{6.7}$ ratio of the entire sample 
varies between 3 and 30, in a range
of a factor of 10.
The ratio increases further at highest luminosities, 
consistent with the results by  
Klaas et al. (2001) and Elbaz et al. (2002).
If the total IR luminosity $L_{3-1000 \mu m}$ is considered 
instead of $L_{FIR}$, the range 
of $L_{3-1000 \mu m} / L_{6.7}$ 
will be even larger. 

Comparing the FIR properties with optical equivalent widths, 
the IRAS detected sources have the same $EW$ [O\,II] versus $EW$ Ca\,II\,K 
distribution as the IRAS non-detections and do not show a preference for a 
high $EW$ [O\,II] (Fig.\ref{fig_OII_CaII}). 
In particular, for those sources with low $EW$ [O\,II],
the cool dust temperature and the signatures of old stellar populations 
suggests that the FIR luminosity is powered substantially by 
the interstellar radiation field from the old stars rather than purely
by the ongoing starformation. 
If this is the case, then $L_{FIR}$ will not properly measure the ongoing
starforming activity.

To summarise, most (17/19) of the IRAS 60-100\,$\mu$m detected sources
are exceptionally luminous cool dust emitters, but not all of them exhibit 
strong starforming signatures as traced by, for instance,
their [O\,II] equivalent widths. 

\subsection{Optical to millimetre SEDs}

\begin{figure}
  \includegraphics[angle=0,width=\columnwidth,clip=true]{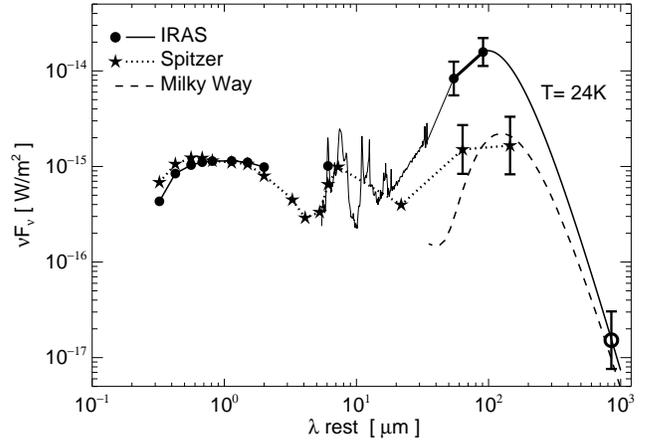}
  \caption{
    Mean optical to millimetre SEDs ($\nu$\,$\cdot$\,$F_{\nu}$) 
    of two subsamples. 
    The upper FIR data points (filled circles) 
    are the IRAS detections;     
    the two ULIRGs have been excluded. 
    The lower FIR photometry curve (asterisks) 
    refers to the Spitzer
    FIR detections. 
    The mean F6.7 of the IRAS sample is a factor of 
    1.5 higher than for the Spitzer sample.  
    The IRS spectrum is 
    that of 2MASS\,0357 (Fig.\,\ref{fig_irs_of_isocp}), scaled to the
    mean F6.7 of the IRAS sample. 
    The optical and NIR data are from SDSS and 2MASS.
    The modified greybody with  temperature $T=24\,K$ 
    has an emissivity index $\beta=2$.  
    The open circle marks the predicted 850\,$\mu$m flux 
    adopting the F6.7\,/\,F850 relation (Haas et al. 2002).
    The long-dashed line shows Milky Way cirrus of the solar
    neighbourhood measured with
    COBE/DIRBE; it is scaled to the mean Spitzer 6.7\,$\mu$m flux level.  
    \label{fig_spitzer_seds}
  }
\end{figure}

Figure~\ref{fig_spitzer_seds} shows the mean optical to millimetre
SEDs ($\nu$\,$\cdot$\,$F_{\nu}$) of the 17 IRAS sources 
and of the 7 Spitzer sources, detected at FIR wavelengths. 
The two ULIRGs have been excluded. 
The important features of these IRAS- and Spitzer-SEDs are:
 
\begin{itemize}
\item[1)]

The Spitzer-SED has three peaks of roughly the same height, 
one in the optical at about 0.6\,$\mu$m, one in the MIR 
around 7\,$\mu$m and one in the FIR at about 100\,$\mu$m. 
Between these peaks the SED shows valleys at about 4 and 25\,$\mu$m.
This SED is similar to that of NGC\,891 and NGC\,6946.  

\item[2)]
The IRAS-SED shows peaks of about the same height at optical and 
MIR wavelengths,
but is about 7-10 times brighter in the FIR than the Spitzer-SED. 
At 6.7\,$\mu$m the IRAS-SED is about a factor of 1.5 higher than the
Spitzer-SED. 
In the optical-NIR the IRAS-SED is somewhat redder than the
Spitzer-SED.\footnote{Because the number of our Spitzer FIR detected 
sources covered by SDSS is very small, 
we used the mean SDSS SED of all non-IRAS detections.
The result is basically the same, but smoother.} 
The redder optical SED is consistent with the IRAS detection statistics: 
While among the 44 sources with red NIR SED (i.e., with $H-K>0.5$ 
in Fig.~\ref{fig_sample_selection})  about one quarter is
detected with IRAS, the fraction of 
the IRAS detections among the remaining 76 bluer sources 
is about a factor of 3 lower. 
Thus, optical selection criteria are biased against the red
but exceptionally FIR bright sources. 

\item[3)]
Fig.~\ref{fig_spitzer_seds} also shows the mean 850\,$\mu$m 
flux predicted for the IRAS sources. 
This prediction is based on the 1:1 
relation between $LW2$ and SCUBA 850\,$\mu$m flux 
found for normal, luminous and
ultra-luminous dust-rich local galaxies (Klaas et
al. 2001, Haas et al. 2002). 
The remarkable benefit of this relation is that the 
sub-mm flux can be predicted from the PAH flux, despite the fact that 
the emission from cold dust is not related to that from
transiently heated very small grains. 
The 60-850 $\mu$m SED is consistent with cool dust 
at $T \la 25\,K$ (emissivity index $\beta=2$). 
Here, we applied a
$K$-correction for average flux loss in $LW2$. Without such a
$K$-correction the dust temperature would be warmer. 
Using standard formulae (Kr\"ugel 2003), high 
dust and gas masses are derived  
($M_{\rm dust} = 10^{8} - 10^{9} M_{\odot}$,  and 
$M_{\rm gas} = 10^{10} - 10^{11} M_{\odot}$ 
assuming the standard gas/dust mass ratio of 100).  
For comparison, the dust masses of M\,31 (Andromeda galaxy) and the
archetypal ULIRG Arp\,220 are about $3\times 10^{7}$ and 
$1.4\times10^{9} M_{\odot}$, respectively (Haas et al. 1998, 
Klaas et al. 2001).  

\item[4)]
For comparison, the Milky Way SED from COBE/DIRBE is shown.
While the peaks at 6.7 and 100\,$\mu$m are of similar height, the
MIR 20-60\,$\mu$m energy output is much lower than in the IRAS- and 
Spitzer-SEDs, and at optical wavelengths the Milky Way would be about ten 
times brighter than in the FIR (not shown in
Fig.~\ref{fig_spitzer_seds} to avoid confusion by too many lines). 
The SEDs of sources with an IRAS FIR upper limit have about equally high 
peaks at optical and 6.7\,$\mu$m. 
This adds a further argument supporting that these sources are similar
to the Spitzer-SED sources, i.e. local templates like NGC\,891 and NGC\,6946. 

\end{itemize}
To summarise, the 
most striking feature of the optical to millimetre SEDs is the 
strong FIR bump in about 15\% (17/120) of the sources. 
Despite being caused by rather cold dust, the FIR energy output of these
sources is about a factor of ten 
higher than at optical wavelengths.  
Future surveys, for instance with the Herschel Space Telescope, 
will find more such galaxies, also at higher redshifts, 
and thus may solve the debate on
how much a cool galaxy population contributes to the 
cosmic infrared background 
(e.g. Lagache et al.  2005, Dole et al. 2006, Juvela et al. 2009).  


\section{Summary and conclusions}

In order to obtain a census of the typical properties of PAH 
emitting galaxies, we combined the ISOCAM Parallel Survey at 
6.7 $\mu$m with 2MASS 
and colour-selected a sample of 120 candidates for strong PAH emission.
Optical spectra of 75 sources establish that they are starforming 
galaxies at moderate distance ($0.03<z<0.3$) with a median redshift 
$z \sim 0.1$. 
Mid-infrared spectroscopy of the reddest 19 sources confirms   
that they have strong PAH emission and that they are not
dust-enshrouded AGN. This leads us to conclude that the 
entire sample of 120 sources consists of PAH selected galaxies 
with a 6.7\,$\mu$m luminosity between 10$^{\rm 8}$
and 10$^{\rm 11}$ $L_{\odot}$. 
This galaxy population has the following properties:

\begin{itemize}

\item [1)]
In about one third of the sources the H$_{\alpha}$ and  
[O\,II]$_{3727\AA}$ equivalent widths indicate intermediate 
to strong starforming activity, while 
one third shows
only moderate ongoing star formation but 
H$_{\delta}$ and/or Ca\,II\,K absorption, indicating a significant 
contribution of stars of intermediate mass and age. 
The remaining third of the sources lies between these cases. 
Many sources exhibit morphological irregularities, but our data 
did not allow us to 
establish a relation with spectral properties. 

\item [2)]
The mid-infrared emission lines [S\,III]$_{\rm ~18.7 \mu m}$ and
[Ne\,III]$_{\rm ~15.5 \mu m}$ as well as the lack of higher ionisation lines 
corroborate that even in the reddest 
sources the radiation field is relatively soft. 
For all sources except two ULIRGs,
the MIR lines and continua exclude powerful buried AGN and 
do not suggest the presence of intense hidden starbursts.
The 24$\mu$m/6.7$\mu$m flux ratio is more like  that of cool normal spiral
galaxies such as NGC\,891 and NGC\,6946 and in a few sources it reaches
that of the Antennae galaxy Arp\,244. 

\item [3)]
In the far-infrared, 26 sources are 
detected on IRAS-ADDSCANS or Spitzer maps. 
Most SEDs steeply rise from 60 to 100\,$\mu$m, adding further
evidence that a large amount of cool dust ($T < 25 K$, 
$M_{dust} > 10^{8} M_{\odot}$) dominates the FIR luminosity 
(10$^{\rm 9}$ to 2\,$\cdot$ \,10$^{\rm 12}$ $L_{\odot}$). 
In 17 sources ($\sim$15\%) the FIR energy output is 5--10
times higher than the optical one, despite the cool 
dust temperature. 
The sources with upper FIR flux limits are
consistent with this picture. 
The exceptional FIR-luminous galaxies increase the 
dispersion in the $L_{PAH} / L_{FIR}$ relation.
Their, on average, redder optical colours and the fact that 30\% of 
them have rather low H$_{\alpha}$ and [O\,II]$_{3727\AA}$ equivalent widths, 
strongly suggest that optical selection criteria fail to collect 
a complete PAH emitting galaxy sample. 

\end{itemize} 
At moderate distance, PAH selected galaxies 
turn out to be a quite heterogeneous population of dust-rich, 
partly infrared-luminous galaxies 
but mostly cool with a range of 
post-starburst signatures and starforming activity 
which appears rather modest relative to the entire gas content. 
The diversity of PAH emitting galaxies strongly suggests that  
the cosmological interpretation of deep 15 and 24\,$\mu$m surveys in
favour of an early universe full of IR starburst galaxies with high
starforming efficiency appears to be premature.



\begin{acknowledgements}
The ISOCAM Parallel Survey has been performed 
with the Infrared Space Observatory
ISO, an ESA project with instruments
funded by ESA Member States  and
with the participation of ISAS and NASA.
The Two Micron All
Sky Survey (2MASS) is a joint project of the University of
Massachusetts and IPAC/Caltech, funded by the National
Aeronautics and Space Administration and the National Science
Foundation. This work is based essentially on observations made
with the Spitzer Space Telescope, which is operated by the
Jet Propulsion Laboratory, California Institute of
Technology under a contract with NASA.
Observing time for optical spectroscopy has been granted at the
telescopes:
South
African Astrophysical Observatory 1.9m, Tautenburg 2m, Kitt Peak 2.1m,
Calar Alto 2.2m, Nordic Optical Telescope 2.5m, Telescopio Nazionale
Galileo 3.5m,
ESO/NTT 3.5m, CTIO Blanco 4m, ESO/VLT 8.2m.
We thank an anonymous referee for detailed constructive comments.
This work was supported by
Nordrhein-Westf\"alische Akademie der Wissenschaften und der K\"unste.
\end{acknowledgements}

\begin{table*}[!h]
\caption{
The sample of PAH selected galaxies and observed properties. 
 }
\label{table_sample}
\scriptsize
\begin{tabular}{l|ccc|rr|c|rrrrrrr|c}
\hline 
2MASS              & LW2    & $H$   & $K_s$ & F$_{60}$  & F$_{100}$&  $z$   & H$\alpha$& H$\beta$ & [O\,II] & EW H$\alpha$& EW [O\,II] &EW H$\delta$& EW Ca\,II\,K&morph  \\ 
                   & mJy a) & mag   & mag   & mJy f)     & mJy    &       &     b)    &     b)   &   b)  &  $\AA$ c)   &  $\AA$ c)& $\AA$ d) &  $\AA$ d)&  e) \\ 
\hline                                                                                                                                                                               
00112015$-$1156099 &   0.98 & 15.81 & 15.47 &   $<$144 &          & 0.241 &      2.33 &     0.78 &          & $-$53.19 &          &     3.75 &          &   \\ 
00180953$-$1009448 &   0.77 & 16.01 & 15.10 &   $<$129 &          &       &           &          &          &          &          &          &          &  compact \\ 
00253965$-$3252466 &   2.58 & 15.41 & 15.19 &   $<$147 &          &       &           &          &          &          &          &          &          &   \\ 
00285431$-$7726146 &   3.45 & 15.28 & 14.78 &      190 &      590 & 0.087 &     10.76 &     0.91 &     2.11 & $-$46.98 & $-$31.13 &     5.22 &     3.46 &   \\ 
00445167$-$2653191 &   1.73 & 15.51 & 14.73 &   $<$132 &          & 0.113 &      9.52 &     0.70 &     2.06 & $-$18.42 &  $-$8.06 &     5.48 &     4.21 &   \\ 
00465933$-$2527218 &   1.37 & 15.71 & 15.50 &   $<$123 &          &       &           &          &          &          &          &          &          &   \\ 
00482481$-$2519270 &   0.95 & 15.34 & 15.41 &   $<$138 &          &       &           &          &          &          &          &          &          &   \\ 
00524246$-$3024175 &   1.51 & 15.40 & 14.90 &   $<$129 &          &       &           &          &          &          &          &          &          &   \\ 
00544355$+$0024382 &   1.99 & 15.88 & 15.83 &   $<$132 &          &       &           &          &          &          &          &          &          &  edge-on \\ 
00544465$-$3907006 &   3.12 & 15.90 & 15.44 &   $<$135 &          &       &           &          &          &          &          &          &          &   \\ 
00561930$-$3742185 &   1.08 & 16.10 & 15.21 &    $<$96 &          & 0.090 &      2.20 &          &          &  $-$7.50 &          &     3.80 &          &   \\ 
00565066$-$2736010 &   1.05 & 16.36 & 15.78 &   $<$114 &          & 0.109 &      5.80 &     0.72 &     1.44 & $-$45.90 & $-$14.10 &     2.20 &          &   \\ 
01051501$-$2612466 &   1.71 & 16.15 & 15.43 &   $<$132 &          & 0.110 &     11.97 &     2.08 &     4.14 & $-$94.84 & $-$40.77 &     2.86 &     2.01 &   \\ 
01144234$-$4521493 &   0.96 & 16.17 & 15.38 &    $<$96 &          & 0.123 &      9.31 &     0.85 &          & $-$54.94 &          &          &          &   \\ 
01240481$+$0403022 &   1.57 & 15.90 & 15.33 &   $<$120 &          & 0.110 &      9.00 &     0.93 &          & $-$42.00 &          &     2.60 &          &   \\ 
01255879$+$8739093 &   2.13 & 15.93 & 15.55 &   $<$162 &          &       &           &          &          &          &          &          &          &   \\ 
01511176$-$0931553 &   0.80 & 15.86 & 14.92 &   $<$150 &          &       &           &          &          &          &          &          &          &  compact \\ 
01515168$+$0253146 &   1.40 & 15.67 & 15.11 &   $<$123 &          & 0.128 &      8.67 &     0.75 &     1.40 & $-$32.19 &  $-$9.50 &     1.50 &     7.79 &   \\ 
02072273$+$2338418 &   2.04 & 13.74 & 13.62 &    $<$81 &          &       &           &          &          &          &          &          &          &   \\ 
02185851$-$0242460 &   1.53 & 15.51 & 15.26 &   $<$156 &          &       &           &          &          &          &          &          &          &   \\ 
02261823$-$1510595 &   1.40 & 15.56 & 15.19 &   $<$111 &          & 0.070 &     38.43 &     5.01 &          & $-$35.91 &          &     2.00 &          &   \\ 
02542961$+$1509122 &   1.36 & 16.14 & 14.92 &   $<$147 &          & 0.099 &     25.43 &     1.31 &          & $-$40.30 &          &     6.00 &     6.61 &   \\ 
03262563$+$3039153 &   2.97 & 14.73 & 14.60 &   $<$138 &          &       &           &          &          &          &          &          &          &   \\ 
03280979$+$3126028 &   1.54 & 14.23 & 13.90 &   $<$210 &          &       &           &          &          &          &          &          &          &   \\ 
03525104$+$1120254 &   1.04 & 15.39 & 14.77 &   $<$156 &          &       &           &          &          &          &          &          &          &  compact \\ 
03561911$-$6423202 &   1.03 & 15.82 & 15.33 &          &          & 0.059 &     34.12 &     9.24 &          & $-$78.06 &          &     2.00 &          &   \\ 
03574895$-$1340458 &  10.67 & 14.52 & 13.87 &      160 &      620 & 0.072 &     42.70 &     3.76 &          & $-$54.02 &          &     3.00 &     2.40 &   \\ 
04145870$+$0547135 &   1.91 & 15.49 & 15.09 &    $<$96 &          & 0.096 &      4.44 &     0.84 &     1.38 & $-$85.28 & $-$49.82 &     2.00 &          &   \\ 
05375494$-$4408547 &   1.47 & 15.48 & 15.04 &    $<$78 &          & 0.105 &      1.41 &     0.14 &     0.22 & $-$24.12 &  $-$9.27 &     2.96 &     5.87 &   \\ 
05400261$-$6002387 &   2.26 & 15.38 & 14.39 &      210 &      430 & 0.155 &     12.00 &          &     3.61 & $-$30.00 & $-$20.00 &     4.30 &     6.00 &   \\ 
05575775$-$3808188 &   1.15 & 15.15 & 14.81 &    $<$60 &          & 0.102 &     14.00 &          &          & $-$34.00 &          &          &          &   \\ 
05580218$-$3805258 &   0.88 & 15.57 & 14.91 &    $<$87 &          & 0.033 &      9.50 &          &          & $-$12.80 &          &          &          &   \\ 
06033357$-$4509412\parbox{0cm}{$^{\rm ~u}$} &   3.51 & 15.28 & 14.19 &     1680 &     2300 & 0.161 &     16.00 &  $-$1.14 &     4.46 & $-$48.00 & $-$46.49 &     5.48 &     1.90 &   \\ 
06213825$+$7813105 &   1.76 & 14.71 & 14.19 &    $<$99 &          &       &           &          &          &          &          &          &          &   \\ 
08083959$+$7554555 &   0.76 & 15.51 & 14.89 &   $<$114 &          &       &           &          &          &          &          &          &          &   \\ 
11110523$+$5550119 &   1.17 & 15.82 & 14.98 &   $<$108 &          & 0.141 &     11.01 &     2.10 &     2.35 & $-$59.09 & $-$18.28 &     2.10 &     3.42 &   \\ 
12191852$+$2953166 &   3.06 & 14.99 & 14.46 &   $<$213 &          & 0.104 &     10.09 &          &     1.70 & $-$16.11 &  $-$9.30 &          &          &  tail \\ 
12211572$+$1151156 &   0.99 & 15.58 & 15.18 &   $<$117 &          &       &           &          &          &          &          &          &          &  group \\ 
12255713$+$4838051 &   0.83 & 16.31 & 15.49 &   $<$174 &          & 0.261 &      7.86 &     1.84 &          & $-$41.30 &          &          &          &  irregular \\ 
12281094$+$0852471 &   2.29 & 15.43 & 15.07 &   $<$114 &          & 0.088 &     53.90 &    12.40 &    28.00 & $-$50.77 & $-$39.20 &     1.80 &     2.60 &  regular \\ 
12325569$+$1328585 &   1.93 & 16.23 & 15.99 &   $<$111 &          &       &           &          &          &          &          &          &          &  companion? \\ 
12361826$-$3938464 &   1.53 & 15.84 & 15.10 &   $<$138 &          & 0.105 &      8.17 &     1.99 &          & $-$27.74 &          &          &          &   \\ 
12361865$-$3938006 &   3.19 & 15.56 & 15.10 &      230 &          & 0.109 &      1.64 &     0.17 &     0.23 & $-$43.24 & $-$29.62 &     2.19 &     4.20 &   \\ 
12482307$-$0344102 &   1.87 & 15.26 & 14.55 &      200 &      470 & 0.176 &     10.98 &     0.72 &     2.55 & $-$23.80 & $-$14.01 &     5.80 &     4.94 &  regular \\ 
13092436$+$2939351 &   1.26 & 16.23 & 15.51 &   $<$168 &          & 0.113 &      9.64 &     0.80 &     7.70 & $-$33.56 & $-$15.00 &          &          &  companion? \\ 
13165999$-$3336459 &   3.53 & 15.00 & 14.29 &      120 &          & 0.111 &      4.19 &     0.15 &     0.42 & $-$13.75 &  $-$7.06 &     3.70 &     5.56 &   \\ 
13371372$+$0906127 &   1.41 & 15.84 & 15.33 &   $<$150 &          & 0.138 &     26.96 &     4.35 &    13.01 & $-$35.34 & $-$29.00 &     2.10 &     2.29 &  edge-on \\ 
13371721$+$0904430\parbox{0cm}{$^{\rm ~u}$} &   1.45 & 16.10 & 14.93 &      140 &      780 & 0.306 &     17.03 &     1.89 &     7.20 & $-$52.67 & $-$45.70 &     4.50 &     2.55 &  companion \\ 
13371806$-$3005123 &   0.92 & 15.99 & 15.59 &   $<$156 &          & 0.037 &     20.90 &     6.20 &          & $-$23.50 &          &          &          &   \\ 
13484308$+$4912271 &   1.45 & 15.44 & 15.02 &   $<$120 &          & 0.136 &      5.74 &     1.10 &     1.48 & $-$53.61 & $-$15.51 &     3.80 &     3.80 &  compact \\ 
13491845$+$1808258 &   1.07 & 15.53 & 14.94 &   $<$108 &          & 0.086 &     25.60 &          &          & $-$18.50 &          &          &          &  regular \\ 
\hline  
\end{tabular}
\end{table*}

\addtocounter{table}{-1}
\begin{table*}
 \caption[]{ continued. }
\label{msxxxx_tab_fluxes_a} 
\scriptsize
\begin{tabular}{l|ccc|rr|c|rrrrrrr|c}
\hline 
2MASS              & LW2    & $H$   & $K_s$ & F$_{60}$  & F$_{100}$&  $z$    &  H$\alpha$& H$\beta$ & [O\,II] & EW H$\alpha$& EW [O\,II] &EW H$\delta$& EW Ca\,II\,K & morph \\ 
                   & mJy a) & mag   & mag   & mJy f)   & mJy      &         &  b)    &     b)   &   b)  &  $\AA$ c)   &  $\AA$ c)& $\AA$ d) &  $\AA$ d)&  e) \\ 
\hline                                                                                                                                                                        
14025863$+$6426228 &   0.98 & 16.17 & 15.71 &    $<$99 &          & 0.122 &     14.70 &     1.95 &     5.50 & $-$28.12 & $-$26.45 &     3.00 &     2.20 &  regular \\ 
14080590$+$5414370 &   0.82 & 15.89 & 15.56 &    $<$78 &          &       &           &          &          &          &          &          &          &  tail \\ 
14234395$+$2552515 &   1.71 & 15.66 & 15.31 &   $<$162 &          & 0.150 &      7.50 &     1.40 &          & $-$28.80 &          &          &          &  compact \\ 
14335047$+$1037077 &   1.04 & 15.52 & 15.07 &   $<$117 &          &   0.109 &  28.50 &     4.67 &     6.78 & $-$26.68 & $-$10.16 &     2.50 &     5.70 &regular  \\ 
14343833$+$3051499 &   0.95 & 15.77 & 14.86 &    $<$96 &          &         &        &          &          &          &          &          &          &double? \\ 
14375074$-$1442498 &   1.41 & 16.05 & 15.51 &   $<$114 &          &   0.216 &   5.12 &  $-$0.50 &     1.00 & $-$29.95 & $-$12.00 &     2.50 &     7.18 &  \\ 
14404891$-$0011195 &   1.65 & 15.95 & 15.51 &   $<$138 &          &   0.072 &  16.08 &     3.62 &     6.73 & $-$22.42 & $-$16.50 &     1.70 &     3.20 &regular  \\ 
14445354$+$2919055 &   2.97 & 14.46 & 14.07 &    $<$90 &          &         &        &          &          &          &          &          &          &double?  \\ 
14511245$+$2253010 &   2.56 & 15.21 & 14.56 &      250 &      330 &   0.119 &   9.90 &          &          & $-$36.75 &          &          &          &edge-on  \\ 
14523534$+$7354420 &   1.57 & 15.50 & 14.91 &    $<$87 &          &   0.176 &   9.08 &     0.90 &     4.30 & $-$35.03 & $-$20.00 &     3.10 &          &  \\ 
14535143$+$2431031 &   1.33 & 15.82 & 14.94 &    $<$90 &          &   0.065 &   1.30 &     0.30 &          & $-$27.88 &          &          &          &edge-on  \\ 
15004675$+$4803272 &   2.08 & 15.70 & 15.27 &      290 &      610 &   0.140 &  17.40 &     2.88 &     3.94 & $-$22.00 & $-$10.10 &     2.00 &     5.50 &regular  \\ 
15125997$+$0724457 &   2.51 & 15.08 & 14.72 &      210 &          &   0.101 &  94.19 &    17.04 &    17.02 & $-$44.55 & $-$12.43 &     3.50 &     5.80 &regular  \\ 
15160785$-$1608098 &   1.54 & 15.69 & 14.94 &      210 &      810 &   0.129 &   8.41 &     1.18 &     1.27 & $-$33.14 & $-$25.10 &     2.50 &     4.20 &  \\ 
15163000$+$5614003 &   0.81 & 15.97 & 15.81 &   $<$195 &          &         &        &          &          &          &          &          &          &compact  \\ 
15270421$+$1928206 &   1.39 & 15.53 & 14.82 &   $<$132 &          &   0.121 &   2.20 &     0.20 &     0.30 & $-$54.74 & $-$13.90 &     2.00 &          &companion?  \\ 
15492435$+$4822553 &   0.76 & 15.68 & 15.47 &    $<$87 &          &   0.072 &  54.60 &    13.60 &    37.60 & $-$54.65 & $-$38.38 &     3.60 &     1.40 &compact  \\ 
15554606$+$1532218 &   2.06 & 15.69 & 14.70 &   $<$111 &          &   0.131 &  33.01 &     8.90 &    19.09 & $-$66.00 & $-$43.00 &     1.70 &     2.00 &tail  \\ 
16025967$+$5826592 &   2.01 & 15.35 & 14.95 &      150 &          &         &        &          &          &          &          &          &          &  \\ 
16053289$+$2037347 &   2.45 & 15.55 & 15.17 &    $<$75 &          &   0.133 &  89.67 &    18.97 &    38.55 & $-$69.73 & $-$34.86 &     3.00 &     2.36 &companion?  \\ 
16132299$+$4203418 &   1.99 & 16.29 & 15.12 &      210 &      310 &   0.138 &  13.09 &     3.41 &     8.41 & $-$64.43 & $-$34.90 &     2.50 &     3.66 &irregular  \\ 
16164591$+$3221543 &   1.77 & 15.84 & 15.52 &      130 &      430 &   0.103 &  48.00 &    10.96 &    23.50 & $-$65.00 & $-$29.33 &     3.80 &     2.40 &regular  \\ 
16251436$+$2352427 &   1.44 & 15.65 & 15.16 &   $<$123 &          &   0.063 &  18.80 &     3.15 &     7.85 & $-$27.38 & $-$18.64 &     2.60 &     3.20 &regular  \\ 
16275436$+$5533101 &   1.17 & 16.07 & 15.46 &   $<$126 &          &   0.135 &  22.26 &     1.32 &          & $-$33.28 &          &          &          &regular  \\ 
16452008$+$4615126 &   1.12 & 15.66 & 15.28 &   $<$108 &          &         &        &          &          &          &          &          &          &compact  \\ 
17193516$+$4751517 &   0.96 & 15.74 & 15.23 &   $<$147 &          &   0.118 &  15.08 &     2.63 &     3.31 & $-$26.93 & $-$17.05 &     4.74 &     5.00 &  \\ 
17350268$+$6814020 &   0.78 & 16.00 & 15.51 &    $<$63 &          &   0.029 &   1.60 &     0.31 &          & $-$18.34 &          &          &          &  \\ 
17480224$+$6723023 &   1.42 & 15.64 & 15.41 &    $<$84 &          &         &        &          &          &          &          &          &          &  \\ 
17493662$+$6832182 &   0.79 & 15.69 & 14.96 &    $<$87 &          &         &        &          &          &          &          &          &          &  \\ 
17512100$+$5652085 &   0.96 & 16.17 & 15.67 &    $<$81 &          &   0.154 &   3.17 &     0.38 &          & $-$33.32 &          &          &          &regular  \\ 
17533157$+$5710232 &   1.50 & 15.30 & 14.75 &   $<$220 &          &   0.068 &   6.51 &          &          & $-$29.86 &          &          &          &companion?  \\ 
17542851$+$5818226 &   0.73 & 15.20 & 15.33 &    $<$98 &          &         &        &          &          &          &          &          &          &compact  \\ 
17545857$+$5121560 &   1.15 & 15.82 & 15.19 &    $<$78 &          &   0.071 &   3.50 &     0.94 &          &  $-$9.70 &          &          &          &regular  \\ 
17554655$+$6638399 &   1.74 & 15.39 & 15.02 &    $<$42 &          &   0.043 &   0.22 &     0.03 &          & $-$64.79 &          &          &          &  \\ 
17562602$+$5136032 &   1.38 & 15.42 & 14.95 &    $<$84 &          &   0.118 &  14.02 &     1.41 &          & $-$65.63 &          &          &          &  \\ 
17564103$+$6634162 &   1.99 & 15.30 & 14.91 &    $<$33 &          &         &        &          &          &          &          &          &          &  \\ 
17571162$+$6645220 &   1.22 & 15.94 & 15.67 &   $<$135 &          &         &        &          &          &          &          &          &          &  \\ 
17574141$+$6648578 &   1.03 & 15.49 & 15.15 &   $<$108 &          &         &        &          &          &          &          &          &          &  \\ 
17582761$+$6651113 &   2.10 & 15.40 & 14.97 &    $<$45 &          &   0.118 &  14.90 &     2.02 &          & $-$25.87 &          &          &     3.66 &  \\ 
17584647$+$6647585 &   1.48 & 15.82 & 15.04 &    $<$75 &          &   0.077 &  46.82 &     5.98 &     7.20 & $-$39.80 & $-$21.95 &     3.30 &          &  \\ 
18143059$+$4512303 &   1.85 & 13.99 & 13.74 &    $<$74 &          &         &        &          &          &          &          &          &          &  \\ 
18233704$+$6418121 &   0.86 & 14.94 & 14.60 &    $<$93 &          &         &        &          &          &          &          &          &          &regular  \\ 
18332203$+$6336009 &   1.79 & 15.30 & 14.74 &    $<$90 &          &   0.083 &  28.39 &     4.78 &     5.15 & $-$63.98 & $-$35.90 &     4.50 &          &regular  \\ 
19135367$+$6732331 &   7.37 & 12.89 & 12.83 &    $<$90 &          &         &        &          &          &          &          &          &          &  \\ 
19184395$-$3815471 &   1.75 & 15.68 & 14.97 &   $<$276 &          &   0.160 &        &     1.19 &     0.92 &          & $-$44.00 &     2.50 &          &  \\ 
19283755$-$4124470 &   1.54 & 14.19 & 14.05 &    $<$69 &          &         &        &          &          &          &          &          &          &  \\ 
20101411$-$0616586 &   4.39 & 15.66 & 15.51 &   $<$150 &          &         &        &          &          &          &          &          &          &  \\ 
20285026$-$1947520 &   1.05 & 16.10 & 15.24 &   $<$108 &          &   0.159 &  11.53 &     1.88 &          & $-$17.77 &          &          &          &  \\ 
20445728$-$1049227 &   0.85 & 14.94 & 14.43 &    $<$75 &          &         &        &          &          &          &          &          &          &  \\ 
\hline  
\end{tabular}
\end{table*}

\addtocounter{table}{-1}
\begin{table*}
 \caption[]{ continued. }
\label{msxxxx_tab_fluxes_b} 
\scriptsize
\begin{tabular}{l|ccc|rr|c|rrrrrrr|c}
\hline 
2MASS              & LW2    & $H$   & $K_s$ & F$_{60}$  & F$_{100}$&  $z$    &  H$\alpha$& H$\beta$ & [O\,II] & EW H$\alpha$& EW [O\,II] &EW H$\delta$& EW Ca\,II\,K & morph \\ 
                   & mJy a) & mag   & mag   & mJy f)   & mJy      &         &  b)    &     b)   &   b)  &  $\AA$ c)   &  $\AA$ c)& $\AA$ d) &  $\AA$ d)&  e) \\ 
\hline                                                                                                                                                                        
21251104$-$1736181 &   4.02 & 15.37 & 14.91 &      150 &          &   0.089 &   9.38 &     1.11 &          & $-$79.67 &          &          &          &  \\ 
21573517$+$0114331 &   1.83 & 15.90 & 15.21 &      200 &          &   0.136 &  41.32 &     8.73 &     9.67 & $-$56.55 & $-$15.52 &     3.80 &     2.87 &regular  \\ 
21582549$-$3022493 &   0.82 & 15.94 & 15.38 &   $<$153 &          &   0.175 &        &  $-$0.67 &     0.44 &          &  $-$7.57 &     5.06 &     7.21 &  \\ 
22023161$-$5657577 &   2.29 & 15.90 & 15.23 &    $<$99 &          &   0.083 &  17.43 &     4.44 &    13.57 & $-$33.50 & $-$16.65 &     4.54 &     4.80 &  \\ 
22064236$-$0353402 &   2.68 & 15.33 & 14.71 &      180 &      520 &   0.072 &  19.37 &     0.56 &          & $-$30.74 &          &          &          &  \\ 
22352396$-$2649402 &   1.36 & 15.44 & 15.17 &   $<$108 &          &         &        &          &          &          &          &          &          &  \\ 
22485447$-$5134173 &   1.42 & 15.76 & 14.89 &      110 &      660 &   0.094 &  14.44 &     2.08 &     9.99 & $-$44.48 & $-$30.45 &     2.00 &          &  \\ 
22564496$-$3637423 &   1.04 & 15.56 & 14.82 &   $<$156 &          &         &        &          &          &          &          &          &          &  \\ 
22565483$-$3649515 &   1.27 & 16.05 & 15.21 &   $<$156 &          &   0.086 &   6.72 &     1.17 &     3.16 & $-$29.40 & $-$15.33 &     4.99 &     3.27 &  \\ 
22572343$-$2950575 &   0.99 & 15.96 & 15.03 &    $<$90 &          &         &        &          &          &          &          &          &          &  \\ 
23032475$+$1652437 &   0.82 & 15.99 & 15.06 &   $<$123 &          &         &        &          &          &          &          &          &          &  \\ 
23055422$-$3539591 &   1.22 & 15.78 & 14.98 &    $<$87 &          &   0.162 &   9.13 &     1.16 &     1.60 & $-$25.69 & $-$11.34 &     4.27 &     6.46 &  \\ 
23075850$+$0537063 &   1.21 & 15.83 & 15.13 &   $<$126 &          &   0.136 &   8.50 &     1.90 &          & $-$19.48 &          &          &          &  \\ 
23080919$+$0538305 &   1.47 & 16.01 & 14.82 &   $<$129 &          &   0.146 &  27.40 &     5.53 &          & $-$21.05 & $-$12.35 &          &          &  \\ 
23094110$-$2236359 &   1.58 & 15.03 & 14.39 &          &          &         &        &          &          &          &          &          &          &  \\ 
23134646$-$6252115 &   1.11 & 15.46 & 15.30 &   $<$102 &          &         &        &          &          &          &          &          &          &  \\ 
23242107$-$0555516 &   3.76 & 15.08 & 14.93 &   $<$144 &          &         &        &          &          &          &          &          &          &  \\ 
23345423$+$4025496 &   1.17 & 15.38 & 14.90 &   $<$140 &          &         &        &          &          &          &          &          &          &  \\ 
23370573$-$0208169 &   2.25 & 15.61 & 15.12 &   $<$165 &          &   0.078 &   9.37 &     1.61 &          & $-$67.09 &          &          &          &  \\ 
23521660$+$2000553 &   1.37 & 14.94 & 15.05 &   $<$111 &          &         &        &          &          &          &          &          &          &  \\ 
\hline  
\end{tabular}
~\\ %
a) LW2 zero magnitude corresponds to 90 Jy, i.e. 0.9 mJy corresponds to 12.5 mag (Vega system). \\ %
b) Emission line fluxes as measured in the 1-2$\arcsec$ wide slit are given in 10$^{\rm -16}$ erg/s/cm$^{2}$/$\AA$, the uncertainties are $\sim$ 5 - 10\%. \\ %
c) Negative EW values correspond to emission lines. Mean uncertainties are $\sim$ 10\%. \\ %
d) Mean uncertainties are $\sim$ 25\%. \\ %
e) Morphological appearance in SDSS images. The definitions are: 
{\it regular}, {\it edge-on} and {\it irregular} sources are extended
with $>5\arcsec$ diameter, regular sources have between face-on and
intermediate inclination;  
{\it compact}   = regular and $<5\arcsec$ diameter; 
{\it tail }     = irregular, but with bright connected tidal tail(s); 
{\it double}    = two equally bright sources; 
{\it companion} = another fainter source nearby ($<10\arcsec$); 
{\it group}     = at least two companions. \\ %
f) Two sources are not covered by IRAS. \\ %
u) ULIRG \\ %
\end{table*}    

\begin{table}
\caption{Parameters from the Spitzer IRS spectra.
The second column gives the fraction of the 6-10\,$\mu$m flux seen 
in the $LW2$ passband (5-8.5\,$\mu$m). 
F15 refers to the flux integrated over the ISOCAM 
$LW3$ passband (12-18\,$\mu$m). 
The uncertainty of the continua F15, F24, and F35 are about 10\%.
Emission line fluxes are given in 10$^{\rm -15}$ erg/s/cm$^{2}$ with $\sim$ 10-20\,\% uncertainties.
}
\label{table_mir_continuum}
\scriptsize
\renewcommand{\footnoterule}{}  
\begin{tabular}{l|r|rrr|rr}
\hline 
2MASS      & LW2 /        & F15  &  F24  &  F35  &[Ne\,III]&[S\,III]  \\
           & 6-10\,$\mu$m & mJy  &  mJy  &  mJy  & 15.55   & 18.71    \\
\hline                                          
00285431$-$7726146                        & 0.57 & 7.03 &   9.3 &  27.2 &    6.1 &  16.0 \\ 
00445167$-$2653191                        &      &      &   4.8 &  11.7 &    -   &   -   \\ 
01051501$-$2612466                        & 0.49 & 2.88 &   4.6 &  11.6 &    5.8 &  12.0 \\ 
02542961$+$1509122                        & 0.52 & 2.15 &   3.0 &   8.1 &    1.6 &   4.2 \\ 
03574895$-$1340458                        & 0.61 &14.48 &  25.2 &  85.0 &    3.6 &  19.1 \\ 
06033357$-$4509412                        &      &      &  89.3 & 468.5 &    -   &   -   \\ 
12482307$-$0344102                        &      &      &   4.7 &  12.0 &    -   &   2.4 \\ 
13165999$-$3336459                        &      &      &   5.3 &  14.2 &    -   &   2.6 \\ 
13371721$+$0904430                        &      &      &   6.2 &  22.8 &    -   &   2.1 \\ 
15160785$-$1608098                        &      &      &   3.3 &   9.5 &    -   &   3.0 \\ 
15554606$+$1532218\parbox{0cm}{$^{\rm ~*}$} & 0.40 & 2.88 &   5.3 &  13.7 &    2.5 &   5.9 \\ 
16132299$+$4203418                        &      &      &  13.6 &  51.4 &    -   &   6.1 \\ 
19184395$-$3815471                        &      &      &   7.7 &  24.8 &    -   &   0.8 \\ 
21573517$+$0114331                        & 0.42 & 3.06 &   3.9 &   9.6 & $<$0.7 &   3.2 \\ 
22023161$-$5657577                        & 0.58 & 4.34 &   7.6 &  18.7 & $<$2.0 &   8.1 \\ 
22565483$-$3649515                        & 0.58 & 1.60 &   2.2 &   6.0 & $<$1.0 &   5.2 \\ 
23055422$-$3539591                        &      &      &   3.0 &   7.7 &    -   &   -   \\ 
23075850$+$0537063                        &      &      &   3.2 &   9.5 &    -   &   -   \\ 
23080919$+$0538305                        & 0.41 & 3.22 &   4.6 &  11.6 & $<$0.7 &   4.1 \\ 
\hline
\end{tabular}
~\\
$*$)  absolute photometry uncertain, see footnote in Section~\ref{sect_data_mir_spectroscopy}.

\end{table}    

\begin{table}
\caption{Photometry from Spitzer IRAC (3.6-8\,$\mu$m) and MIPS
  (24-160\,$\mu$m). 
  Uncertainties are about 10\% at 3.6-8\,$\mu$m,  and less than 30\% 
  at 24-160\,$\mu$m.
}
\label{table_spitzer_photometry}
\scriptsize
\renewcommand{\footnoterule}{}  
\begin{tabular}{l|rrrrrrr}
\hline
2MASS &  F3.6  &    F4.5  &    F5.8  &    F8.0  &    F24 &     F70  &    F160\\
      &   mJy  &    mJy   &    mJy   &    mJy   &   mJy  &    mJy   &    mJy \\
\hline
00465933$-$2527218\parbox{0cm}{$^{\rm ~*}$} &        &          &          &          &    5.29&     53.1 &   158.0\\ 
00482481$-$2519270\parbox{0cm}{$^{\rm ~*}$} &        &          &          &          &    2.52&     28.5 &    81.0\\ 
00561930$-$3742185                        &  0.477 &          &    0.410 &          &    1.50&     21.5 &    49.9\\ 
13092436$+$2939351                        &        &    0.530 &          &    3.09  &        &          &        \\ 
13371806$-$3005123                        &  0.676 &          &    0.898 &          &    1.57&     21.4 &        \\ 
14080590$+$5414370\parbox{0cm}{$^{\rm ~*}$} &        &          &          &          &    6.57&     55.4 &        \\ 
15163000$+$5614003\parbox{0cm}{$^{\rm ~*}$} &  0.335 &    0.309 &    0.290 &    1.61  &    2.75&     53.9 &    76.3\\ 
17512100$+$5652085                        &        &          &          &          &    2.21&          &        \\ 
17554655$+$6638399                        &  0.520 &    0.38  &    0.86  &    2.96  &    4.49&          &        \\ 
17564103$+$6634162                        &  1.050 &    0.714 &    0.915 &    4.37  &    4.87&          &        \\ 
17571162$+$6645220\parbox{0cm}{$^{\rm ~*}$} &  0.455 &    0.337 &    0.645 &    1.66  &        &     34.6 &        \\ 
17574141$+$6648578                        &  0.572 &    0.410 &    0.557 &    1.91  &    1.40&          &        \\ 
17582761$+$6651113                        &  0.705 &    0.520 &    0.630 &    3.39  &    2.65&          &        \\ 
17584647$+$6647585                        &  0.527 &    0.375 &    0.576 &    2.57  &    2.28&          &        \\ 
\hline
\end{tabular}
~\\
$*$)  Median redshift of the sample ($z=0.105$) adopted.
\end{table}    


\begin{thebibliography}{}



\bibitem[2008]{bendo08} 
  Bendo, G. J.; Draine, B. T.; Engelbracht,
  C. W. et al. 2008, MNRAS 389, 629

\bibitem[2008]{bregman08} 
  Bregman, J. D.; Bregman, J. N.; Temi, P., 2008, ASPC 381, 34

\bibitem[2001]{calzetti01} 
  Calzetti, D. 2001, PASP 113, 1449

\bibitem[2007]{caputi07} 
  Caputi, K. I.; Lagache, G.; Yan, L.; et al. 2007, ApJ 660, 97


\bibitem[1996]{cesarsky_c96} 
  Cesarsky, C. J., Abergel, A., Agnese, P.; et al. 1996, A\&A  315, L32

\bibitem[1996]{cesarsky_d96} 
  Cesarsky, D.; Lequeux, J.; Abergel, A.; et al. 1996, A\&A 315, L309

\bibitem[2002]{chapman02} 
  Chapman S.C., Smail I., Ivison R.J et al. 2002, ApJ 573, 66

\bibitem[1986]{chini86} 
  Chini, R.; Kreysa, E.; Kruegel, E.; Mezger, P. G. 1986, A\&A 166, L8 

\bibitem[2001]{dale01} 
  Dale, D. A.; Helou, G.; Contursi, A.; et al. 2001, ApJ 549, 215

\bibitem[2005]{dale05} 
  Dale, D. A.; Bendo, G. J.; Engelbracht, C. W.;
  et al. 2005, ApJ 633, 857


\bibitem[2005]{dennefeld05}
  Dennefeld, M.; Lagache, G.; Mei, S.; et. al. 2005, A\&A 440, 5

\bibitem[2006]{dole06}
  Dole, H.; Lagache, G.; Puget, J.-L.; et al. 2006, A\&A 451, 417

\bibitem[1982]{dressler92}
  Dressler, A.; Gunn, J. E. 1982, ApJ 263, 533


\bibitem[2002]{elbaz02}
  Elbaz, D.; Cesarsky, C. J.; Chanial, P.; et al. 2002, A\&A 384, 848

\bibitem[1998]{genzel98} Genzel, R., Lutz, D., Sturm, E., et al. 1998, ApJ,
  498, 579

\bibitem[1998]{haas98}
  Haas, M.; Lemke, D.; Stickel, M.; et al. 1998, A\&A 338, L33

\bibitem[2000]{haas00}
  Haas, M.; Klaas, U.; Coulson, I.; et al. 2000, A\&A 356, L83


\bibitem[2002]{haas02} Haas, M., Klaas U., Bianchi S. 
  2002, A\&A  325, L23

\bibitem[2004]{haas04} Haas, M., Siebenmorgen, R., Leipski, C., et
  al. 2004, A\&A  419, L49



\bibitem[2004]{houck04} 
  Houck, J. R.; Roellig, T. L.; van Cleve, J.; et al. 2004, ApJS 154, 18

\bibitem[2009]{juvela09} Juvela M., Mattila K., Lemke D., et al., 
  2009, A\&A ~in press, 2009arXiv0904.2997J

\bibitem[2008]{kaneda08}
  Kaneda, H.; Onaka, T.; Sakon, I.; et al. 2008, ApJ 684, 270

\bibitem[1992]{kennicutt92}
  Kennicutt, R. C. 1992, ApJ 388, 310

\bibitem[2001]{klaas01}
  Klaas U.; Haas M.; M\"uller, S. A. H.; et al. 2001, A\&A 379, 823

\bibitem[2003]{kruegel03} 
  Kr\"ugel E., 2003 'The physics of interstellar dust',
  IoP Series in Astron. \& Astrophys. Bristol


\bibitem[1996]{kunze96}
  Kunze, D.; Rigopoulou, D.; Lutz, D.; et al. 1996, A\&A 315, L101

\bibitem[2005]{lagache05} 
  Lagache, G.; Puget, J.-L.; Dole, H. 2005, ARA\&A 43, 727

\bibitem[2000]{laurent00}
  Laurent, O.; Mirabel, I. F.; Charmandaris, V.; et al. 2000, A\&A 359,
  887 


\bibitem[2005]{leipski05} Leipski, C., Haas, M., Meusinger, H., et al.
  2005, A\&A  440, L8

\bibitem[2007]{leipski07} 
  Leipski, C.; Haas, M.; Meusinger, H.; et al.
  2007, A\&A  464, 895


\bibitem[1998]{lemke98}
  Lemke, D.; Mattila, K.; Lehtinen, K.; et al. A\&A 331, 742

\bibitem[1995]{liu95}
  Liu, C. T.; Kennicutt, R. C. 1995, ApJ 450, 547


\bibitem[1996]{mattila96} Mattila, K.; Lemke, D.; Haikala, L. K.; et al.
  1996, A\&A  315, L353

\bibitem[1999]{mattila99} Mattila, K.; Lehtinen, K.; Lemke, D.
  1999, A\&A  342, 643

\bibitem[1998]{mirabel98} 
  Mirabel, I. F.; Vigroux, L.; Charmandaris, V.; et al. 1998, A\&A
  333, L1

\bibitem[2008]{monkiewicz08}
  Monkiewicz, J. A.; Dickinson, M. E.; Davoodi, P.; et al. 2008, ASPC
  381, 332


\bibitem[2003]{ott03} Ott, S., Siebenmorgen, R., Schartel, N., V\~o, T.
  2003, ESA SP-511, 159

\bibitem[2003]{patris03}
Patris, J.; Dennefeld, M.; Lagache, G.; Dole, H. 2003, A\&A 412, 349

\bibitem[1989]{puget89}
  Puget, J. L.; L\'eger, A. 1989, ARA\&A 27, 161


\bibitem[2004]{mrr04}
  Rowan-Robinson, M.; Lari, C.; Perez-Fournon, I.; et al. 2004, MNRAS
  315, 1290

\bibitem[2006]{sajina06}
  Sajina, A.; Scott, D.; Dennefeld, M et al. 2006, MNRAS 369, 939


\bibitem[1996]{sanders96}
  Sanders, D. B.; Mirabel, I. F. 1996, ARA\&A 34, 749

\bibitem[2006]{schweitzer06}
  Schweitzer, M.; Lutz, D.; Sturm, E.; et al. 2006, ApJ 649, 79 

\bibitem[2007]{shi07}
  Shi, Y.; Ogle, P.; Rieke, G. H.; et al. 2007, ApJ 669, 841 

\bibitem[1996]{siebenmorgen96} Siebenmorgen, R., Abergel, A., Altieri, B.,
  et al. 1996, A\&A  315, L169

\bibitem[2000]{siebenmorgen00} 
  Siebenmorgen, R., Schartel, N., \&
  Ott, S. 2000, LNP 548, 275 
  in {\it ISO Surveys  of a Dusty Universe}, ed. D. Lemke et al., 275




\bibitem[1999]{sodre99}
  Sodr\'e, L.; Stasi\'nska, G. 1999, A\&A 345, 391

\bibitem[1987]{thronson87}
  Thronson, H. A., Jr.; Walker, C. K.; Walker, C. E.; Maloney, P. 1987,
  ApJ 318, 645

\bibitem[2000]{uchida00}
  Uchida, K. I.; Sellgren, K.; Werner, M. W.; Houdashelt, M. L. 2000, ApJ
  530, 817


\bibitem[2002]{vaisanen02}
  V\"ais\"anen, P.; Morel, T.; Rowan-Robinson, M. et al. 2002, 
  MNRAS 337, 1043

\bibitem[1987]{veilleux87}
  Veilleux S.; Osterbrock D.E. 1987, ApJS 63, 295

\bibitem[2003]{verma03}
  Verma, A.; Lutz, D.; Sturm, E.;  et al. 2003, A\&A 403, 829


\bibitem[2004]{werner04}
  Werner, M. W.; Roellig, T. L.; Low, F. J.; et al. 2004, ApJS 154, 1

\bibitem[2005]{Wu2005}
  Wu, H.; Cao, C.; Hao, C.-Na; et al. 2005, ApJ 632, L79 

\bibitem[2000]{xu2000}
  Xu, C. 2000 ApJ 541, 134

\bibitem[]{}
  Zhu, Yi-N.; Wu, H.; Cao, C.; Li, H.-N. 2008, ApJ 686, 155


\end{thebibliography}
\end{document}